\documentclass[sigconf,screen,dvipsnames]{acmart} 
\usepackage[T1]{fontenc}
\usepackage{xcolor}
\usepackage{graphics}
\usepackage{subcaption}
\usepackage{booktabs}
\usepackage{enumitem}

\usepackage{amsmath}
\DeclareMathOperator*{\argmax}{arg\,max}
\DeclareMathOperator*{\argmin}{arg\,min}

\newtheorem{definition}{Definition}

\newcommand{\dmSet}[1]{\mathcal{#1}}

\newcommand{\users}[0]{\dmSet{U}}
\newcommand{\user}[0]{u}
\newcommand{\items}[0]{\dmSet{I}}

\newcommand{\observedItems}[0]{\tilde{\items}}

\newcommand{\quality}[0]{\sigma}

\newcommand{\rating}[0]{r}
\newcommand{\target}[0]{\rating}
\newcommand{\observed}[0]{\hat{\rating}}
\newcommand{\prediction}[0]{\tilde{\rating}}

\copyrightyear{2020}
\acmYear{2020}
\setcopyright{acmlicensed}\acmConference[SIGIR '20]{Proceedings of the 43rd International ACM SIGIR Conference on Research and Development in Information Retrieval}{July 25--30, 2020}{Virtual Event, China}
\acmBooktitle{Proceedings of the 43rd International ACM SIGIR Conference on Research and Development in Information Retrieval (SIGIR '20), July 25--30, 2020, Virtual Event, China}
\acmPrice{15.00}
\acmDOI{10.1145/3397271.3401034}
\acmISBN{978-1-4503-8016-4/20/07}

\begin{document}


\title[Operationalizing the Legal Principle of Data Minimization for Personalization]{Operationalizing the Legal Principle of Data Minimization\\ for Personalization}

%
\author{Asia J. Biega}
\affiliation{%
  \institution{Microsoft Research}
  \city{Montr\'eal}
}
\author{Peter Potash}
\affiliation{%
  \institution{Microsoft Research}
  \city{Montr\'eal}
}
\author{Hal Daum\'e III}
\affiliation{%
  \institution{Microsoft Research NYC}
}
\affiliation{
\institution{University of Maryland}
}
\author{Fernando Diaz}
\affiliation{%
  \institution{Microsoft Research}
  \city{Montr\'eal}
}
\author{Mich\`ele Finck}
\affiliation{%
  \institution{Max Planck Institute for Innovation and Competition}
}

%

\newcommand{\abcomment}[1]{\textcolor{BlueGreen}{{\bf [AB: }{\em #1}{\bf ]}}}

\begin{abstract}
Article 5(1)(c) of the European Union's General Data Protection Regulation (GDPR) requires that "personal data shall be [...] adequate, relevant, and limited to what is necessary in relation to the purposes for which they are processed (`data minimisation')". To date, the legal and computational definitions of `purpose limitation' and `data minimization' remain largely unclear. In particular, the interpretation of these principles is an open issue for information access systems that optimize for user experience through personalization and do not strictly require personal data collection for the delivery of basic service.

In this paper, we identify a lack of a homogeneous interpretation of the data minimization principle and explore two operational definitions applicable in the context of personalization. The focus of our empirical study in the domain of recommender systems is on providing foundational insights about the (i) feasibility of different data minimization definitions, (ii) robustness of different recommendation algorithms to minimization, and (iii) performance of different minimization strategies.We find that the performance decrease incurred by data minimization might not be substantial, but that it might disparately impact different users---a finding which has implications for the viability of different formal minimization definitions. Overall, our analysis uncovers the complexities of the data minimization problem in the context of personalization and maps the remaining computational and regulatory challenges.
\end{abstract}

%
%

%

%

%
\maketitle

\section{Introduction}
\label{sec:introduction}
%
%
Personalized services such as recommender systems or search engines collect large amounts of user interaction logs. 
Such data collection practice is widely accepted to be necessary for platforms to build high-quality models  \cite{halevy:unreasonable-data,sun:unreasonable-data}.
%
%
However, some prior work shows that exact user interaction profiles are not necessary to tailor the results of search or recommendations. For instance, 
Singla \emph{et al.}
show that it is possible to personalize results while storing a reduced user interaction history \cite{singla2014stochastic}, while 
Biega \emph{et al.}
show that it is possible to shuffle user queries and ratings while preserving the quality of personalized search and recommendations \cite{biega2017privacy}.

%
%
If results can be personalized without exact user profiles, it is pertinent to ask: \textit{How much information and what information does an individual need to provide to receive quality personalized results}? Note the parallel between this question and the principle of \textit{data minimization} defined in Article 5 of the European Union's General Data Protection Regulation (GDPR)~\cite{regulation2016regulation} as well as data protection regimes in other jurisdictions, which requires that a system only retain user data necessary to deliver service. The core idea we explore in this work is whether the principles of purpose limitation and data minimization can be complied with in the context of personalization and what minimizing data in this context entails. 

%
%
%
%
In contrast to other GDPR concepts, such as \textit{the right to be forgotten} or \textit{informed consent}, there is to date only marginal regulatory and judicial guidance on the interpretation of data minimization.
Reasoning about data minimization has largely been confined to setups involving immutable or relatively stationary user characteristics. For instance, examples mentioned in the guidelines issued by the UK's Information Commissioner's Office~\cite{icoDataMinimization}
discuss scenarios of collecting people's names by debt collectors, or employee blood types by employers. More recent regulatory guidelines and industrial practice, however, recognize the multitude of challenges related to minimization in data-intensive applications~\cite{icoDataMinimizationTechniques, galdon2020auditing}.

To the best of our knowledge, this work is the first to operationalize the legal concepts of purpose limitation and data minimization in a scenario where \textit{user data collection is not strictly necessary to deliver a service, but where the collection of such data might improve service quality}. We tie the purpose of data collection to performance metrics, and define \emph{performance-based minimization principles}.

%
%
In this study, we investigate two possible technical definitions of performance-based data minimization. The first interpretation, which we refer to as \textit{global data minimization}, minimizes per-user data collection subject to meeting a target mean performance across users.  This aligns well with standard empirical risk minimization approaches in machine learning~\cite{vapnik1992principles}. Alternatively, \textit{per-user data minimization} minimizes per-user data collection subject to each user meeting a target performance. Equivalently, this aligns with meeting a target performance for the minimum across all users.

We use these interpretations to compare different minimization strategies for personalized recommendations.  We demonstrate that quality recommendations can be provided while collecting substantially less user data.  However, we also find that the appropriate minimization strategy is sensitive to the base recommendation algorithm used.  While our results suggest that systems should be able to achieve global data minimization, we demonstrate that preserving the average performance conceals  substantial impact for individual users. 
%
%
To sum up, the salient contributions of this paper are:
\begin{itemize}[noitemsep,topsep=0pt,parsep=0pt,partopsep=0pt,leftmargin=10pt]
    \item Identifying a lack of a homogeneous interpretation of the GDPR's purpose limitation and data minimization principles in the context of personalization systems and proposing a computational definition of \emph{performance-based data minimization}.
    \item An analysis of the feasibility of two different data minimization definitions in the domain of recommender systems.
    \item An analysis of the robustness of different recommendation algorithms to various minimization strategies, both on a population as well as an individual user levels.
\end{itemize}

\section{Data Minimization}

\subsection{A legal perspective}
Article 5(1)(c) GDPR requires that personal data be \emph{\lq adequate, relevant and limited to what is necessary in relation to the purposes for which they are processed.\rq} Data minimisation is the direct consequence of the legal principle of \emph{purpose limitation}, which requires that personal data only be processed for specified, explicit and legitimate purposes and not further processed in a manner incompatible with these purposes. While these core data protection principles cannot be examined exhaustively here, it is worth noting that general statements such as \lq improving user experience\rq ~are generally not specific enough to meet the legal threshold of purpose limitation. This raises the question of whether \lq personalization\rq can be a purpose under the GDPR at all. 

According to data minimisation, no more personal data than necessary to achieve the purpose can be processed. The first question to ask is thus whether data such as this studied in our paper is personal data. The Article 4 GDPR embraces a very broad definition of personal data as \lq any information relating to an identified or identifiable natural person.\rq In the past, movie ratings such as those in the MovieLens 20M dataset~\cite{harper2016movielens} have been shown to allow for identification through linking of private and public datasets~\cite{narayanan2008robust}. It is thus safe to assume that much of the data used in recommender systems, such as movie ratings, constitutes personal data and is hence subject to the GDPR (where within its geographical scope). 

Data minimisation can be broken down into three distinct requirements. 
First, data must be \emph{adequate} in relation to the purpose that is pursued. 
Arguably, adequacy is the most interesting of the three criteria as it may actually (and somewhat counterintuitively) require that more data is processed. It is well established that the omission of certain data can limit the usefulness of a dataset and the accuracy of an analysis done on that dataset. As such, to achieve accurate results, more data may need to be collected. 
Data minimisation indeed ought to be interpreted in light of the other substantive requirements in Article 5 GDPR such as fairness, transparency and accuracy and there are scenarios, often those involving underrepresented groups, where this can only be achieved through the processing of more personal data. 

Second, data ought to be \emph{relevant} in light of the purpose, meaning that only data that is pertinent for the purpose can be processed. For example, if an e-commerce provider requested users' date of birth to provide personalised recommendations regarding future purchases, this data is unlikely to be relevant (except where recommendations have an astrological flavor). Relevance thus functions as a safeguard against accumulating data simply for the sake of doing so. 

Third, the GDPR requires that data be \emph{limited} to what is necessary, meaning that controllers ought to identify the minimum amount of personal data required to fulfil the stated purpose. Thus, where similarly robust results can be achieved through the processing of less personal data, the processing of personal data can likely not be accepted as being necessary. Where possible, only anonymised data should be used. However, given the practical limitations of achieving anonymisation, the latter cannot be assumed as a viable alternative to minimisation in many contexts~\cite{finck2019they}.

\subsection{Performance-Based Data Minimization}
Our focus in this paper is on operationalizing the third requirement of data minimization, namely that of limitation. According to the legal considerations detailed in the previous subsection, generic statements such as `improving user experience' are not specific enough to be used as a purpose of data collection. Thus, we propose to reason about data minimization by tying the \emph{purpose} to \emph{performance metrics}.
While there are many ways in which this proposition might be operationalized, in this paper, we begin investigating this space with an empirical study of two definitions.

Let $\users$ be a set of users for whom the system needs to minimize the data and let $\items$ be the set of items that a system can recommend.  Each user has \textit{rated} some subset $\items_\user\subseteq \items$ of items.  Let $\target_\user$ be the $|\items|\times 1$ vector of ratings for these items.  Of the rated items in $\items_\user$, in a minimization setting, the system only sees a subset $\observedItems_\user\subseteq \items_\user$, referred to as the observational pool for user $\user$.  Let $\observed_\user$ be the ratings for these observations.   Given $\observed_\user$, a system generates $\prediction_\user$, its predicted ratings for $\user$.  The quality metric for $\user$ is defined as $\quality(\prediction_\user)$.  

\begin{definition}[Global data minimization]
A system satisfies global data minimization if it minimizes the amount of per-user data while achieving the quality of a system with access to the full data on average,
\vspace{-4px}
\begin{align*}
    \min&\phantom{=}k
    \phantom{=}\text{s.t.}&\phantom{=}\forall \user, |\observedItems_\user| = k &\phantom{=}\text{and}
    &\phantom{=}\mathbb{E}_\users[\quality(\prediction'_\user)]-\mathbb{E}_\users[\quality(\prediction_\user)] \leq \lambda
\vspace{-5px}
\end{align*}
where $\prediction'$ is the prediction using the ratings in $\items_\user$ and $\lambda$ is a threshold difference in the expected per-user performance.
\end{definition}

\begin{definition}[Per-user data minimization]
A system satisfies per-user data minimization if it minimizes the amount of per-user data while achieving the quality of a system with access to the full data for each user,
\vspace{-4px}
\begin{align*}
    \min&\phantom{=}k
    \phantom{=}\text{s.t.}&\phantom{=}\forall \user, |\observedItems_\user| = k
    &\phantom{=}\text{and}
    &\phantom{=}\forall \user, \quality(\prediction'_\user) - \quality(\prediction_\user) \leq \lambda
\vspace{-5px}
\end{align*}
where $\prediction'$ is the prediction using the ratings in $\items_\user$ and $\lambda$ is a threshold difference in the per-user performance. 
\end{definition}

\section{Experimental Setup}
\subsection{Datasets}
\label{sec:dataset}
We run our analyses using (1) the MovieLens 20M dataset \cite{harper2016movielens} and (2) the Google Location dataset~\cite{he2017translation}. Because of the space constraints, we report the results using dataset (1), and use dataset (2) for validation, reporting differences in observations where applicable. To properly reason about data minimization, we only select users who have at least 45 ratings in their profile. For efficiency reasons, we further subsample the users, creating (1) a MovieLens dataset containing around 2.5k users, 170k ratings, and 20k unique movies; the mean and median number of ratings in a user profile are 69.5 and 59, respectively, and (2) a Google Location dataset containing around 2.2k users, 185k ratings, and 150k unique items; the mean and median number of ratings in a user profile are 85.2 and 64, respectively.

\subsection{Recommendation algorithms}
\label{sec:recommendation-algorithms}
We analyze data minimization properties for two fundamental classes of recommendation algorithms - neighborhood-based (k-nearest-neighbors) and matrix-factorization-based (SVD)~\cite{funk2006netflix},
both as implemented in the Surprise library~\cite{Surprise}.

\subsubsection{Notation.} For a user $u$ and item $i$, we use $\target_{ui}$ to denote the true rating given by the user for the item and $\prediction_{ui}$ as the predicted rating by the user for the item from a predictive model,

\subsubsection{Neighborhood-based.}
For the neighborhood-based recommendations, we use the user-user k-nearest-neighbors algorithm setting $k=30$, as per prior studies investigating the recommendation performance in the MovieLens dataset~\cite{ekstrand2012recommenders}.
Rating prediction $\prediction_{ui}$ for user $u$ and item $i$ is computed as a weighted sum of the ratings of $i$ made by $u$'s top-$k$ nearest neighbors among users who rated item $i$: 
\begin{equation}
\label{eq:knn}
\prediction_{ui} = 
\frac{
\sum\limits_{v \in N^k_i(u)} \text{sim}(u, v) \cdot \target_{vi}
}
{
\sum\limits_{v \in N^k_i(u)} \text{sim}(u, v)
}
\end{equation}
where $N^k_i(u)$ is the set of users $v$ who have rated item $i$ and who are most similar to user $u$ by the value of similarity measure $sim(u,v)$.

User similarity is computed as the inverse of the mean squared difference of ratings (with add-1 smoothing) over set $I_u\cap I_v$.

\subsubsection{Matrix-factorization-based.}
For the matrix-factorization-based recommendations, we use an implementation of the FunkSVD algorithm~\cite{funk2006netflix} with 30 latent factors. Rating prediction for user $u$ and item $i$ is computed as:
\begin{equation}
\label{eq:svd-prediction}
\prediction_{ui} = \mu + b_u + b_i + q_i^{\intercal}p_u
\end{equation}
where $q_i$ is a 30-dimensional latent vector representing item $i$, $p_u$ is a 30-dimensional latent vector representing user $u$, $\mu$ is a global mean, and $b_i$ and $b_u$ are item and user biases, respectively.

\subsection{Error measures}
We measure the quality of recommendations using: RMSE (comparing the differences between the predicted and true ratings for all items in the test set and thus assuming a user consumes the whole recommendation set) and NDCG (measuring the quality of the top results with a logarithmic discounting factor for errors in lower ranking positions \cite{NDCG}). 
In our experiments, we set $k=10$.

\subsection{Protocol}
We explore data minimization in the context of a system that begins with extensive data collection for a starting set of users.  This may be gathered in-house or from a representative market not subject to data minimization constraints.  While there will be situations where seed data is unavailable, we leave that for future work. 

To simulate this situation, we randomly split the full dataset into two parts: the system data $D_S$ ($70\%$ of all users), and the minimization data $D_M$ ($30\%$ of all users). Users are randomly assigned to one of these groups. For minimizing users in $D_M$, we further randomly split their ratings into candidate ($70\%$ of all ratings) and test data ($30\%$ of all ratings). Different minimization strategies will select different subsets of each user's candidate data for use by the system.
Recommendations generated based on the selected data from the candidate user data are evaluated using the remaining test data. 

Data is selected from the candidate user data using a chosen minimization strategy and a minimization parameter $n$ (the number of items to select). We run experiments for $n = \{1,3,7,15,100\}$.

\subsection{Data minimization strategies}
\label{sec:minimization-strategies}

When minimizing data, we select a subset of user candidate items to present to the recommendation algorithm. 
While approaches with similar problem structure have used greedy algorithms modeling the information-theoretic utility of data~\cite{krause2010utility}, greedy algorithms are less practical in a data minimization scenario. Since utility of data is tied to a specific recommendation performance metric rather than modeled as information gain, the submodularity and monotonicity properties upon which guarantees on greedy algorithms are based do not necessarily hold. Moreover greedy selection is costly in terms of runtime, since the recommendation algorithm needs to be run for every possible selection.
This section presents the selection strategies we study in this paper.

\subsubsection{Full.}
We compare other minimization strategies against a baseline generating predictions based on full observational pools of users from $D_M$. Formally, $\observedItems_\user= \items_\user$.

\subsubsection{Empirical bounds.}
We compare the minimization results against brute-force baselines that select $1$ item from a user's profile that lead to (i) the highest prediction RMSE (One item worst), (ii) the lowest prediction RMSE (One item best). We also compute (iii) the average RMSE error over all possible 1-item selections (One item avg); this value can be thought of as an empirical expected value of RMSE over 1-item random selections.

\subsubsection{Random.}
This strategy selects $n$ ratings uniformly at random from the observational pools of the minimizing users. The key observation to make here is that this method \emph{will not create random user profiles} as a result, but minimized \emph{average profiles} of each user. That is, if ratings of certain types (e.g., of a certain genre) are common in the full observational profile, they are likely to be preserved through the random sampling.

\subsubsection{Most recent.}
This strategy selects $n$ most recent ratings from the observational pools of the minimizing users. Note that one can expect this method to behave similarly to the random method in case the data is collected over a period of time short enough for the user tastes to stay intact. In case the observational data of each user spans a very long time, we could expect the predictions to be better than random in case the test data is also sampled from the most recent ratings, and worse than random otherwise.

\subsubsection{Most/least favorite.}
This strategies select the $n$ ratings that have the highest/lowest value for a given user, respectively. 

\subsubsection{Most Rated.}
This method uses the system data to determine the selection method.
For a given user, we select the $n$ items that have been rated the most often (by the number of times an item has been rated by all users in the system data).

\subsubsection{Most characteristic.}
This method uses the system data to determine the selection method for a given user. We create binary vector representations of items $b_i$ by allocating each system data user to a dimension of $b_i$ and setting the value to 1 if the user has rated item $i$, and 0 otherwise. We then take the average of all the item representations $b_{avg}$. Finally, for a given user we select the $n$ items with the closest Euclidean distance to the average item representation. Whereas the most rated strategy treats all users the same when creating its counts, this strategy rewards items for being rated by users who have rated many items and penalizes items that have been rated by user who have rated few items.
Formally,
$
    \observedItems_\user= \argmin_{\{i\}} \sum d(b_i,b_{avg})\,\, s.t.\,\, |\{i\}| = n,
$
where $d()$ is the Euclidean distance between two vectors, $b_i$ is the binary representation of item $i$, and $b_{avg}$ is the average item vector; all vectors are computed using the system data. 

\subsubsection{Highest variance.}
This method is based on one of the standard approaches for feature selection in machine learning~\cite{guyon2003introduction}. It uses the system data to determine the selection method for each user by looking at which items have the highest variance in their ratings. 
Formally,
$
    \observedItems_\user= \argmax_{\{i\}} \sum \sigma(\{r_{*i}\})^2\,\, s.t.\,\, |\{i\}| = n,
$
where $\sigma$ is standard deviation, and $\{r_{*i}\}$ is the set of all ratings for item $i$ in the system data.

\section{Global data minimization}
\label{ref:how-much-needed}

To guide the interpretation of the results, we want to make the following remarks. 
Reasoning about feasibility of data minimization, it is important to understand what quality loss we would incur if we based personalized recommendations on minimized user profiles. The main purpose of our experimental study is thus to measure and compare the quality of recommendations under different minimization conditions.

To reason about the efficacy of a minimization condition (maximum size of user profile $n$ and a minimization strategy) for a given recommendation algorithm, we measure the difference in the quality of recommendations obtained under the minimization condition, and the quality of recommendations obtained if the recommendation algorithm sees all available user data (the Full strategy). We conclude that minimization is feasible if this difference is not statistically significant, or if the difference is minimal (low RMSE increase, and low NDCG decrease).

\begin{table}[t]
\tiny
\centering
\caption{Minimization performance for k-NN recommendations macro-averaged over all users. $\ast$ denotes cases when the difference between a given strategy and the 'full' strategy is statistically significant under a two-tailed t-test with p < 0.01 and the Bonferroni correction. Average RMSE and NDCG@10 for non-minimized data is 0.915 and 0.777, respectively. Note that the \emph{lack of statistical significance} suggests a minimization technique is \emph{performing well}.
}
\label{tab:results-all-strategies-knn}
\resizebox{0.47\textwidth}{!}{%
\begin{tabular}{ c  c  c  c  c  c }
\toprule
 & \textbf{n=1} & \textbf{n=3} & \textbf{n=7} & \textbf{n=15} & \textbf{n=100}  \\
\midrule \multicolumn{6}{c}{RMSE} \\ \midrule
\textbf{random} & $ 1.062^*$ & $ 1.051^*$ & $ 1.013^*$ & $ 0.963^*$ & $ 0.915$ \\
\textbf{most recent} & $ 1.044^*$ & $ 1.060^*$ & $ 1.028^*$ & $ 0.974^*$ & $ 0.915$ \\
\textbf{most favorite} & $ 1.053^*$ & $ 1.046^*$ & $ 1.000^*$ & $ 0.957^*$ & $ 0.915$ \\
\textbf{least favorite} & $ 1.049^*$ & $ 1.077^*$ & $ 1.039^*$ & $ 0.983^*$ & $ 0.915$ \\
\textbf{most watched} & $ 1.064^*$ & $ 1.007^*$ & $ 0.966^*$ & $ 0.935^*$ & $ 0.914$ \\
\textbf{most characteristic} & $ 1.008^*$ & $ 1.044^*$ & $ 1.073^*$ & $ 1.024^*$ & $ 0.915$ \\
\textbf{highest variance} & $ 1.055^*$ & $ 1.071^*$ & $ 1.020^*$ & $ 0.955^*$ & $ 0.915$ \\
\midrule \multicolumn{6}{c}{NDCG@10} \\ \midrule
\textbf{random} & $ 0.681^*$ & $ 0.721^*$ & $ 0.743^*$ & $ 0.762^*$ & $ 0.777$ \\
\textbf{most recent} & $ 0.678^*$ & $ 0.708^*$ & $ 0.734^*$ & $ 0.760^*$ & $ 0.777$ \\
\textbf{most favorite} & $ 0.697^*$ & $ 0.730^*$ & $ 0.751^*$ & $ 0.767$ & $ 0.777$ \\
\textbf{least favorite} & $ 0.662^*$ & $ 0.700^*$ & $ 0.733^*$ & $ 0.752^*$ & $ 0.777$ \\
\textbf{most watched} & $ 0.721^*$ & $ 0.746^*$ & $ 0.764^*$ & $ 0.772$ & $ 0.777$ \\
\textbf{most characteristic} & $ 0.637^*$ & $ 0.656^*$ & $ 0.690^*$ & $ 0.737^*$ & $ 0.777$ \\
\textbf{highest variance} & $ 0.664^*$ & $ 0.708^*$ & $ 0.744^*$ & $ 0.766^*$ & $ 0.777$ \\
\bottomrule
\end{tabular}
}
\end{table}

\begin{table}[h]
\tiny
\centering
\caption{Minimization performance for SVD recommendations macro-averaged over all users. $\ast$ denotes cases when the difference between a given strategy and the 'full' strategy is statistically significant under a two-tailed t-test with p < 0.01 and the Bonferroni correction. Average RMSE and NDCG@10 for non-minimized data is 0.818 and 0.793, respectively. Note that the \emph{lack of statistical significance} suggests a minimization technique is \emph{performing well}.}
\label{tab:results-all-strategies-svd}
\resizebox{0.47\textwidth}{!}{%
\begin{tabular}{ c  c  c  c  c  c }
\toprule
 & \textbf{n=1} & \textbf{n=3} & \textbf{n=7} & \textbf{n=15} & \textbf{n=100}  \\
\midrule \multicolumn{6}{c}{RMSE} \\ \midrule
\textbf{random} & $ 0.876^*$ & $ 0.861^*$ & $ 0.843^*$ & $ 0.828^*$ & $ 0.818$ \\
\textbf{most recent} & $ 0.875^*$ & $ 0.864^*$ & $ 0.851^*$ & $ 0.837^*$ & $ 0.820$ \\
\textbf{most favorite} & $ 0.886^*$ & $ 0.913^*$ & $ 0.974^*$ & $ 0.999^*$ & $ 0.820$ \\
\textbf{least favorite} & $ 0.888^*$ & $ 0.934^*$ & $ 1.015^*$ & $ 1.036^*$ & $ 0.824^*$ \\
\textbf{most watched} & $ 0.874^*$ & $ 0.864^*$ & $ 0.849^*$ & $ 0.835^*$ & $ 0.818$ \\
\textbf{most characteristic} & $ 0.873^*$ & $ 0.862^*$ & $ 0.847^*$ & $ 0.837^*$ & $ 0.818$ \\
\textbf{highest variance} & $ 0.874^*$ & $ 0.860^*$ & $ 0.842^*$ & $ 0.830^*$ & $ 0.819$ \\
\midrule \multicolumn{6}{c}{NDCG@10} \\ \midrule
\textbf{random} & $ 0.793$ & $ 0.793$ & $ 0.794$ & $ 0.793$ & $ 0.795$ \\
\textbf{most recent} & $ 0.792$ & $ 0.795$ & $ 0.792$ & $ 0.792$ & $ 0.791$ \\
\textbf{most favorite} & $ 0.793$ & $ 0.794$ & $ 0.793$ & $ 0.794$ & $ 0.791$ \\
\textbf{least favorite} & $ 0.794$ & $ 0.793$ & $ 0.792$ & $ 0.793$ & $ 0.793$ \\
\textbf{most watched} & $ 0.794$ & $ 0.792$ & $ 0.792$ & $ 0.790$ & $ 0.792$ \\
\textbf{most characteristic} & $ 0.794$ & $ 0.792$ & $ 0.793$ & $ 0.793$ & $ 0.794$ \\
\textbf{highest variance} & $ 0.793$ & $ 0.793$ & $ 0.794$ & $ 0.792$ & $ 0.791$ \\
\bottomrule
\end{tabular}
}
\end{table}

 \begin{figure*}[t!p]
       \begin{subfigure}[b]{0.247\textwidth}
               \centering
               \includegraphics[width=0.98\textwidth]{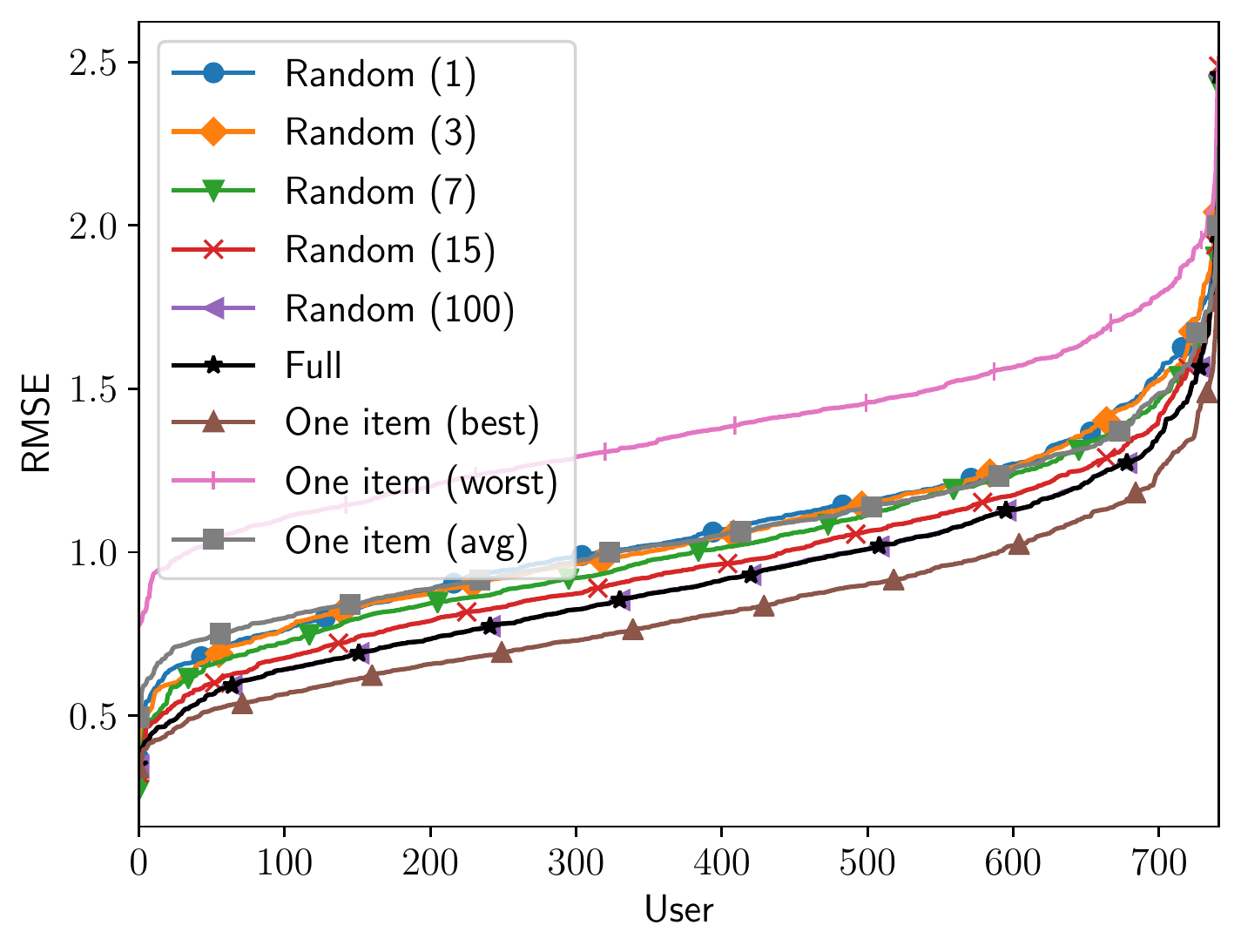}
               \caption{}
               \label{fig:knn-random-rmse-loss-sorted}
     \end{subfigure}
      \begin{subfigure}[b]{0.247\textwidth}
             \centering
             \includegraphics[width=0.98\textwidth]{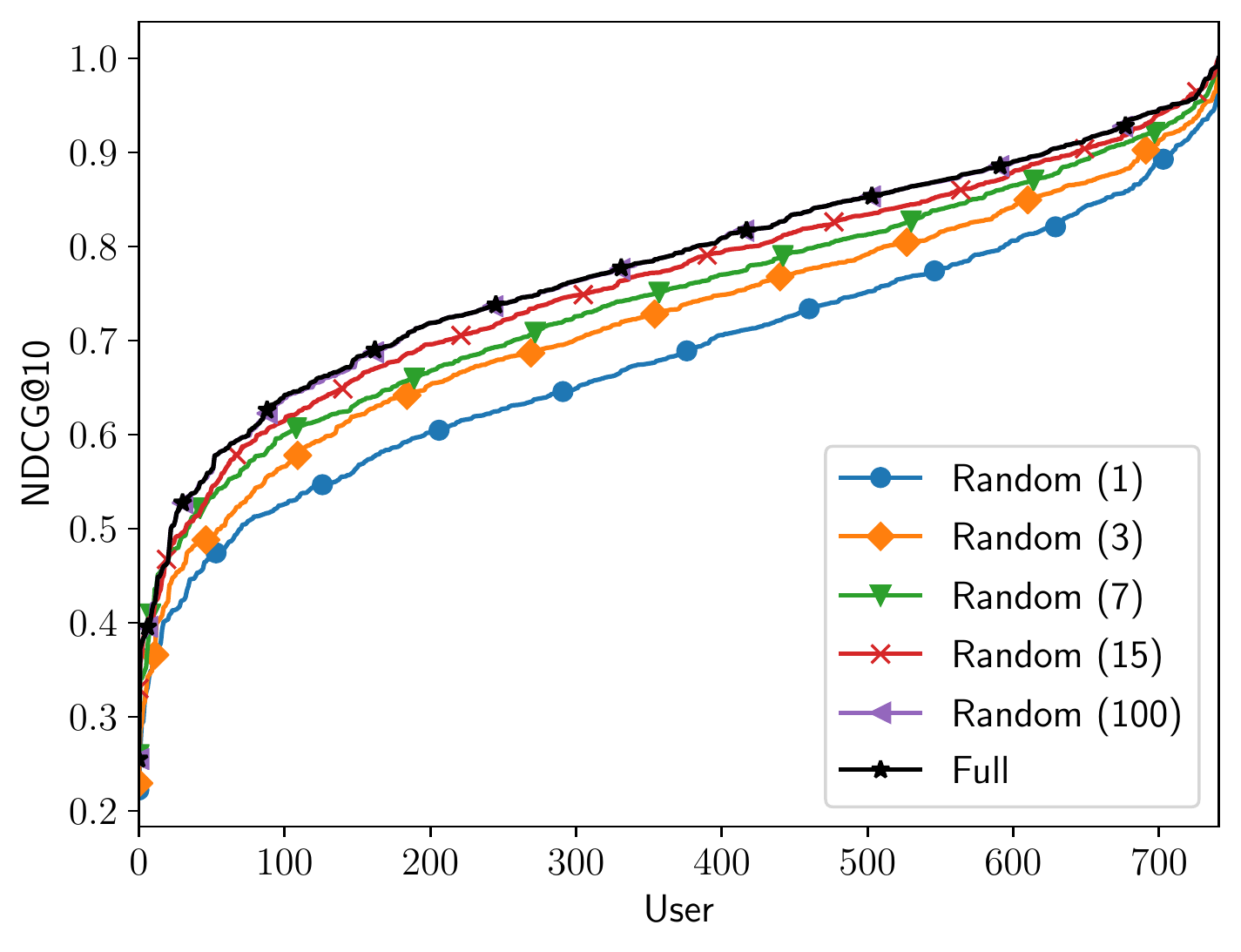}
             \caption{}
             \label{fig:knn-random-ndcg-loss-sorted}
     \end{subfigure}
    \begin{subfigure}[b]{0.247\textwidth}
               \centering
               \includegraphics[width=0.98\textwidth]{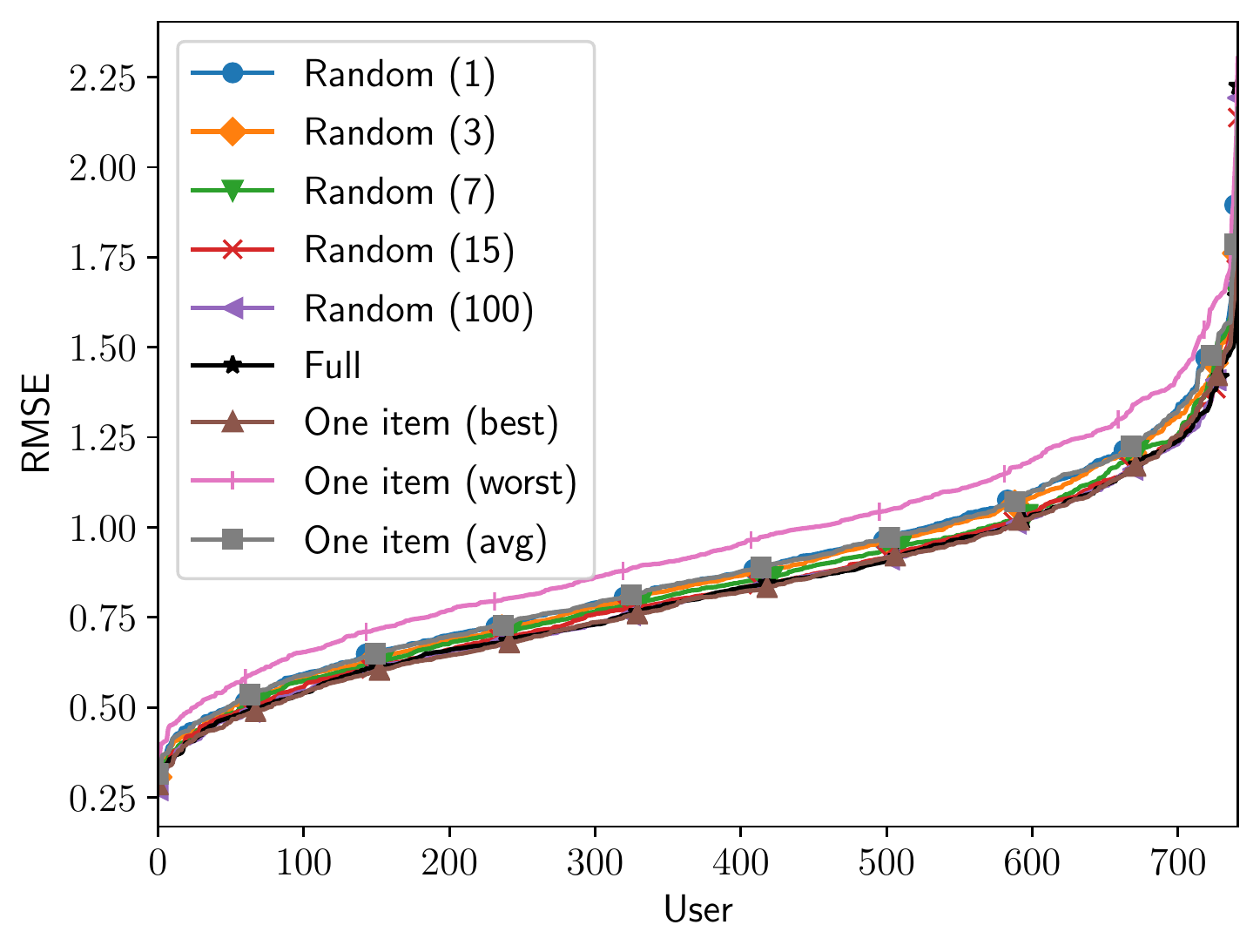}
               \caption{}
               \label{fig:svd-random-rmse-loss-sorted}
     \end{subfigure}
      \begin{subfigure}[b]{0.247\textwidth}
             \centering
             \includegraphics[width=0.98\textwidth]{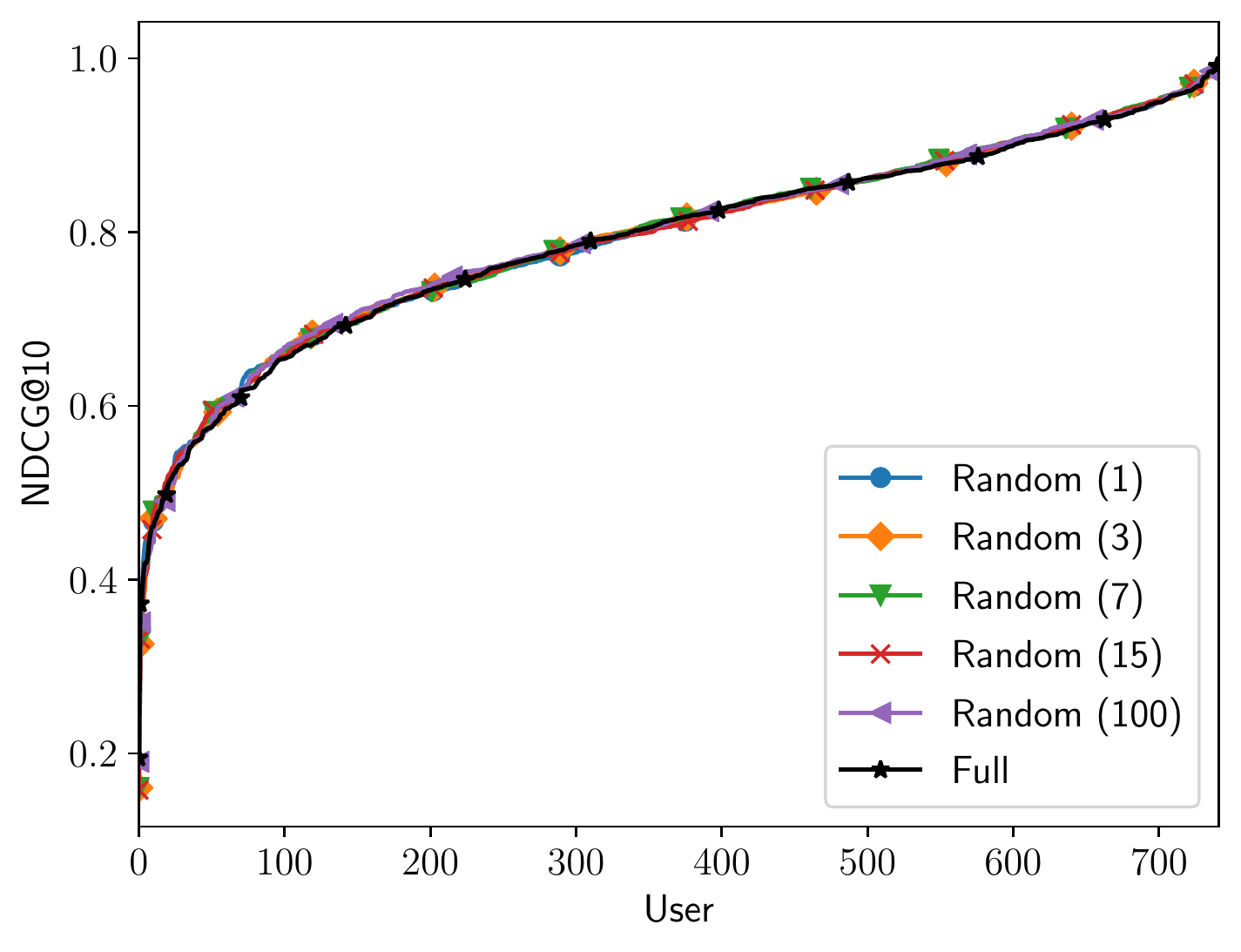}
             \caption{}
             \label{fig:svd-random-ndcg-loss-sorted}
     \end{subfigure}
\caption{Sorted RMSE (a, c) and NDCG (b, d) values for all users when selecting random subsets of items of varying sizes as input to the kNN (a, b) and SVD (c, d) recommendation algorithms. Higher values on the y-axis in plots (a, c) are worse, while higher values on the y-axis in plots (b, d) are better. SVD is more robust to minimization than kNN, with aggressive minimization incurring low quality loss. While error increases as we minimize, the distribution of remains the same.}
\label{fig:global-loss-sorted}
\end{figure*}

\subsection{Feasibility of global data minimization}
Tables~\ref{tab:results-all-strategies-knn} and \ref{tab:results-all-strategies-svd}, as well as Figure~\ref{fig:global-loss-sorted} present the performance of the k-NN and SVD recommendation algorithms for various minimization strategies and intensity (parameter $n$ denotes the number of items from observational pools that were shown to the recommendation algorithm).
The numbers show the RMSE and NDCG values of the minimized recommendations, averaged over all minimizing users. 

For both recommendation algorithms, we observe that the recommendation quality decreases as we present the algorithm with less data to base personalized recommendations on. We attribute the few exceptions (e.g., the increase of RMSE between n=3 and n=7) to the inherent noisiness of data and effects of sampling strategies. 

We would like to highlight two observations.
First, the overall loss incurred by minimization is relatively low when compared to the variation of error across users --- see Figure~\ref{fig:global-loss-sorted} for a visualization of sorted error values for all users in the minimization dataset for random minimization strategies. It is important not to overinterpret these results based on measures like RMSE, though. Ratings in recommendation datasets are often relatively homogenous in terms of absolute values: In the MovieLens dataset, for instance, they vary between $0.5$ and $5$ in $0.5$ increments. Moreover, most users abstain from using extreme values in their ratings: In our system data, out of $10$ different values in the rating scale, the three most used rating values of $3, 3.5,$ and $4$, make $61\%$ of all ratings.

Second, the distribution of error in the population remains the same even when the recommendations are based on minimized data. We observe that the shapes of the error value curves are similar for different minimization methods beyond random (effects similar to those in Figure~\ref{fig:global-loss-sorted}). We exclude additional plots for lack of space.

\subsubsection{Withheld data}
While our experiments explicitly controlled the size of user interaction logs available to a recommendation algorithm, the data \emph{withheld} from the algorithm can be substantial. On average, minimization with $n=\{1,3,7,15,100\}$ leads to $99\%, 96\%, 90\%, 79\%, 6\%$ of data withheld from the recommendation algorithm, respectively. Note that this is not a comment about the total amount of data available to the system: In the setup we consider in this paper, the recommendation algorithm is trained on full data of 70\% of users, which means that the effective percentage of the withheld data is lower. 

\subsection{Algorithm robustness to data minimization}
We find that SVD is more robust to data minimization according to both quality measures. In case of RMSE, metric differences between the Full strategy and any other strategy and minimization parameter $n$ are lower for SVD than for kNN. This observation also holds for NDCG; moreover, the differences in NDCG between the performance of SVD on full data and minimized data are not significant (under a two-tailed t-test and $p<0.01$ with the Bonferroni correction). Note that the SVD robustness result is partly explained by our experimental protocol---the minimized observed data of each test user is 'folded in' into the matrix one user at a time. While this approach is more computationally expensive than folding in all test users at once, the resulting decomposition is computed for a matrix where only one row is different from the full data condition. On top of that, the NDCG measure is not sensitive to differences in predicted rating values as long as the predicted ranking of items remains the same (which is likely to happen when the decomposed matrix is similar to the full data matrix). The lower minimization robustness of kNN can furthermore be explained by the fact that user similarities are computed over rating sets joint with other system users ($I_u\cap I_v$, see~\ref{sec:recommendation-algorithms}), and minimization thus leads to computing predictions over noisier neighbour sets.

\subsubsection{Comparison to prior work.}
Note that these findings are consistent with prior work. First, Chow \textit{et al.}~\cite{chow2013differential} demonstrate that, for similarity-based recommendations, performance often does not differ after removing random data points. Further, different data removal strategies can  improve or degrade predictive performance relative to random removal; in some cases, strategies can improve over the non-minimized predictions \cite[Fig. 1]{chow2013differential}. 

Second, Wen \textit{et al.}~\cite{wen2018exploring} analyzed performance decreases in a recommendation privacy scenario where users provide an algorithm with their recommendation data from the most recent $N$ days. This filtering strategy is similar to the Most Recent minimization strategy we introduce in Sec.~\ref{sec:minimization-strategies}. Wen \textit{et al.} showed that predictions of matrix-factorization-based methods are robust, with performance not degrading even when data is limited to ratings from the previous one to seven days and especially when the percentage of minimizing users is low~\cite[Fig. 2]{wen2018exploring}.\footnote{The experimental protocol used in our paper maps to a setting in Wen et al.~\cite{wen2018exploring} where the percentage $P$ of minimizing users is much lower than $0.25$.}

\subsubsection{Factors influencing rating changes when minimizing for k-NN}
Recall Eq.~\ref{eq:knn}. What will influence the difference between an item prediction $\hat{r}_{ui}$ under the minimization condition and the prediction based on the full observational pool? Since the system data remains intact under our experimental setup, the values of $r_{vi}$ will remain intact as well. The value of $\hat{r}_{ui}$ will be changed, though, when $u$'s relative similarity to other users changes. This might happen when:
\begin{itemize}[noitemsep,topsep=0pt,parsep=0pt,partopsep=0pt,leftmargin=10pt]
    \item The set of nearest neighbors $N^k_i(u)$ changes and user $u$ is placed in a different neighborhood for item $i$. The nearest neighbor summation of $r_{vi}$ ratings happens over a different set of users $v$ (even if the relative similarities to those users stay the same).
    \item The set of nearest neighbors $N^k_i(u)$ changes and user $u$ is placed in a neighborhood where the relative similarities to other users $sim(u,v)$ are different (even if the neighbor rating values $r_{vi}$ are the same).
    \item The set of nearest neighbors $N^k_i(u)$ stays the same but the similarity of $u$ to other users within the neighborhood changes. Note that this is very likely to happen since the similarities will be computed over $u$'s minimized data.
\end{itemize}

While it is possible to enumerate these error contributing factors, analysis of how exactly they impact overall minimization error is challenging because the different dimensions (user similarity, neighborhoods, item popularity, etc.) all influence each other. 

\subsubsection{Factors influencing rating changes when minimizing for SVD}
When will an item prediction $\hat{r}_{ui}$ under the minimization condition and the prediction based on the full observational pool? Note that latent item representations $q_i$ and biases $b_i$ will largely stay intact -- during training, most updates to $q_i$'s and $b_i$'s will come from the data of the system users. The rating change will primarily be influenced by a change in the latent user representation $p_u$ and bias $b_u$ -- during training, updates to these components will come from the latent factors of minimized observational items. Thus, we can expect biggest rating differences if the items in the minimized user profile don't reflect the full user profile.
To examine the relative importance of $p_u$ and $b_u$, we run minimization for recommendations generated using an unbiased SVD (removing $\mu$, $b_u$, and $b_i$ from Eq.~\ref{eq:svd-prediction}). We find that errors incurred by minimization for this setup increase, suggesting that recommendation performance might be preserved by the bias terms when data is minimized.

\subsection{Best and worst minimization strategies}

\subsubsection{Random minimization strategy}
 Figure~\ref{fig:global-loss-sorted} presents sorted RMSE (a, c) and NDCG (b, d) error values per user in the MovieLens dataset population, respectively, when minimizing data using random selection strategies. Unsurprisingly, on average, recommendation error increases as we observe fewer items. The error increase is, however, not substantial. There a number of factors that contribute to this effect. First, note that the random minimization strategy does not create random user profiles, but average user profiles, and the rating distributions over salient categories are likely to remain the same. Second, user profiles are of varying sizes and for some methods minimizing methods already access full observational pools. We tried to alleviate this effect by inclusion of users whose observational pools have at least 45 ratings. To understand these limitations better, we also plot the empirical lower bound on the error for predictions based on empty observational pools (non-personalized predictions based on the system data only). While the random minimization strategy performs reasonably well, there exist better and worse minimization strategies for both recommendation algorithms.
 
\subsubsection{Strategies performing better than random minimization.}
For kNN recommendations, Most Favorite and Most Watched strategies perform better than Random. Movies users like most likely lead to highest contributions to  user-user similarity, and thus the Most Favorite strategy tends to quickly place users in the right neighborhoods. Most Watched, by asking about the most rated movies, will quickly place users belonging to large clusters of popular movie watchers in the right neighborhood. Since there are many users with a taste for most popular movies, this strategy overall leads to a good global minimization performance.

 \subsubsection{Strategies performing worse than random minimization.}
 For kNN recommendations, the Highest Variance selection strategy performs worse than the random selection strategy for the lowest $n$ values ($n=1,3,7$). One hypothesis is that the items selected by this strategy often have very high or very low ratings for a given user, causing this strategy to effectively interpolate between the performance of the Most Favorite and Least Favorite strategies. Whereas Most Favorite usually performs slightly better than random, Least Favorite often performs far worse, and when observed together explains why the Highest Variance strategy often performs worse than the Random selection strategy. 
We believe this is because the most characteristic score is inversely correlated with the most watched count. 

For SVD recommendations, Most Favorite and Least Favorite strategies perform significantly worse than Random. We hypothesize that asking a user for ratings from just one side of their taste spectrum fails to populate all latent dimensions with relevant information. Moreover, since the most and least favorite items of a given user are likely correlated, asking for more items corroborates this effect by constructing an increasingly skewed user taste representations. This skew potentially leads to a reversal effect we have observed---Most Favorite and Least Favorite strategies initially decrease in performance as we increase n.

\subsubsection{Other strategies.}
For kNN recommendations, Most Recent strategy performs on average worse than Random, likely due to the fact that the MovieLens-20M data was collected over a long period of time, yet our testing sample was random. Relatively bad performance of the Least Favorite strategy is related to insensitivity of standard recommendation algorithms to negative rating feedback; systems generally need to be tuned to be able to learn from negative ratings~\cite{frolov2016fifty}.

\subsection{Differences between datasets}
As described in Sec.~\ref{sec:dataset}, we run the the same experiments with two different datsets, using the Google Location dataset for validation. We observe the same trends in terms of the performance of different minimization strategies. One major difference in the results is that we observe similar robustness to minimization for KNN and SVD recommendations.
We attribute this fact to the key difference between the two datasets---in Google Location dataset item ratings are sparser (20k vs. 150k unique items for a similar total number of users) thus minimization is less likely lead to overall change in similarities to other users.

\section{Per-user data minimization}
\subsection{Feasibility of per-user data minimization}
\begin{figure}[t!h!p]
       \begin{subfigure}[b]{0.23\textwidth}
               \centering
               \includegraphics[width=0.98\textwidth]{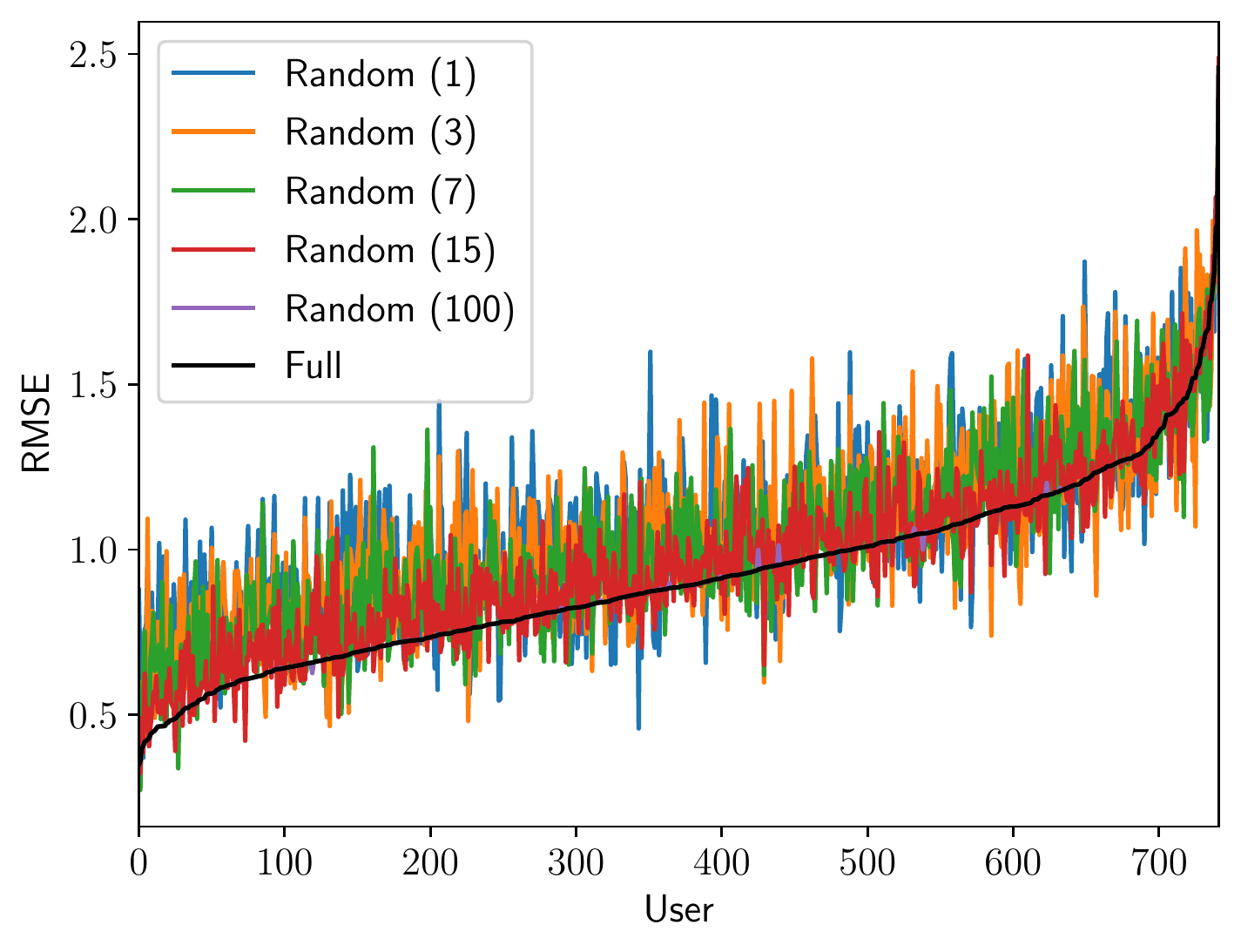}
               \caption{}
               \label{fig:global-rmse-knn-unsorted}
     \end{subfigure}
      \begin{subfigure}[b]{0.23\textwidth}
             \centering
             \includegraphics[width=0.98\textwidth]{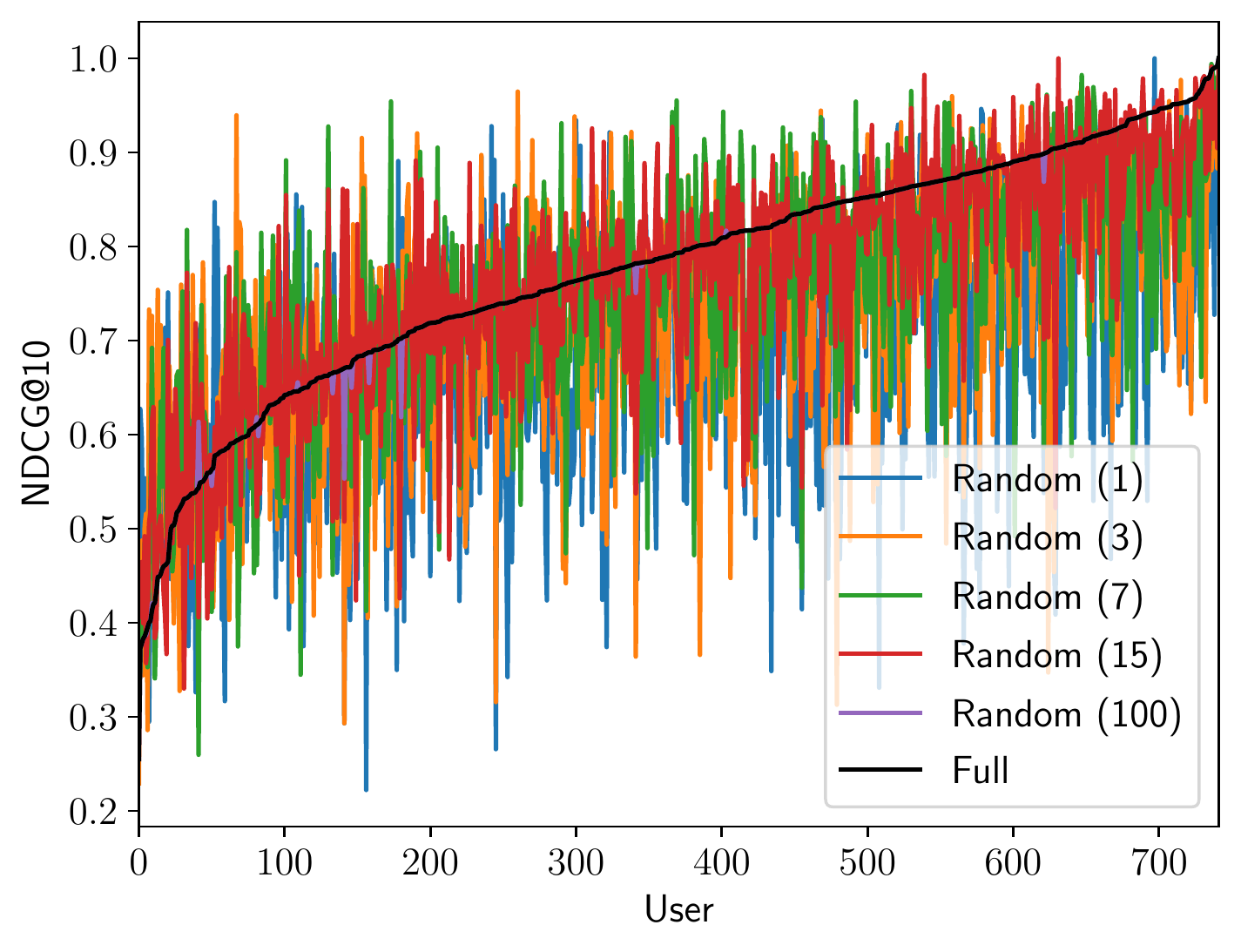}
             \caption{}
             \label{fig:global-ndcg-knn-sorted}
     \end{subfigure}
\caption{RMSE (a) and NDCG (b) variation over the population of users when selecting random subset of items of varying sizes as an input to the kNN algorithm. The underlying data presented here is the same as in Figure \ref{fig:global-loss-sorted}, but the data points are sorted by the y-axis value of the Full strategy only. Data points of other selection methods are unsorted and match the users at the ranking positions defined by the sorting of the Full strategy. This result shows that, while the overall quality loss is low and the error distribution remains the same, the quality loss for individuals can be substantial.}
\label{fig:global-loss-unsorted}
\end{figure}

 Figure~\ref{fig:global-loss-unsorted} shows the error variation when the data is sorted only by the error value of the Full method - other error values correspond to users at the ranking positions determined by the sorting for Full. Note that the data plotted here is exactly the same as the data in Figure~\ref{fig:global-loss-sorted} --- only the sorting differs. 
 These results suggest that, while the distribution of error in the population across users remains largely similar irrespective of recommendation algorithm or minimization strategy (see Figure~\ref{fig:global-loss-sorted}), \emph{errors incurred to individuals can be substantial}. We observe this behavior for all tested minimization methods and recommendation algorithms, although the per-user variations are lower when minimizing for SVD recommendations. 

This finding suggests that, for a fixed quality threshold, data can be less effectively minimized if the loss requirement applies to every individual as opposed to the population on average.

Since the error is not uniformly distributed, we dive deeper to try to understand which users are most impacted. The following sections analyze a number of user characteristics and their correlations with error deltas.

\subsection{User characteristics vs. minimization error}
We investigate whether the minimization errors (the difference in the quality metric when comparing the recommendations over minimized profiles and the recommendations over full profiles) 
are correlated with different user characteristics. For each user, we consider the following characteristics (measured over the user's full profile, before minimization): (1) Number of ratings, (2) Average value of the ratings in a user's profile, (3) Average popularity of items in a user's profile (measured as the number of users in the system data who have rated a given item), (4)  Profile diversity measured by the number of genres the movies in a user's profile belong to, (5) Average similarity to all users in the system data, (6) Average similarity to the 30 most similar users in the system data.

\subsubsection{Regression analysis}
For each pair of recommendation algorithm and minimization strategy, we run an Ordinary Least Squares regression with the error delta as the dependent variable, and the above user characteristics as independent variables. Error delta is computed in two versions as: (i) $\Delta_3=\text{RMSE}(3) - \text{RMSE}(\text{Full})$, and (ii) $\Delta_{15}=\text{RMSE}(15) - \text{RMSE}(\text{Full})$. We compute the coefficient of determination ($R^2$) to measure what proportion of variance in the dependent variable can be explained by the independent variables.

We find that the variance in neither $\Delta_3$ nor $\Delta_{15}$ is well explained by the selected user variables, across recommendation and minimization strategies. For kNN and $\Delta_3$, we get the highest $R^2$ at $0.102$ for the Most Recent strategy, followed by $0.0935$ for the Most Characteristic System strategy, and $0.061$ for the Least Favorite strategy. For kNN and $\Delta_{15}$, $R^2$ values are even lower. For SVD and $\Delta_3$, we get the highest $R^2$ values for the Most and Least Favorite strategies, at $0.396$ and $0.364$, respectively. For SVD and $\Delta_3$, $R^2$ values follow similar trends.

\subsubsection{A closer look}
\begin{figure}[htb]
\centering
      \begin{subfigure}[t]{0.2\textwidth}
             \includegraphics[width=0.98\textwidth]{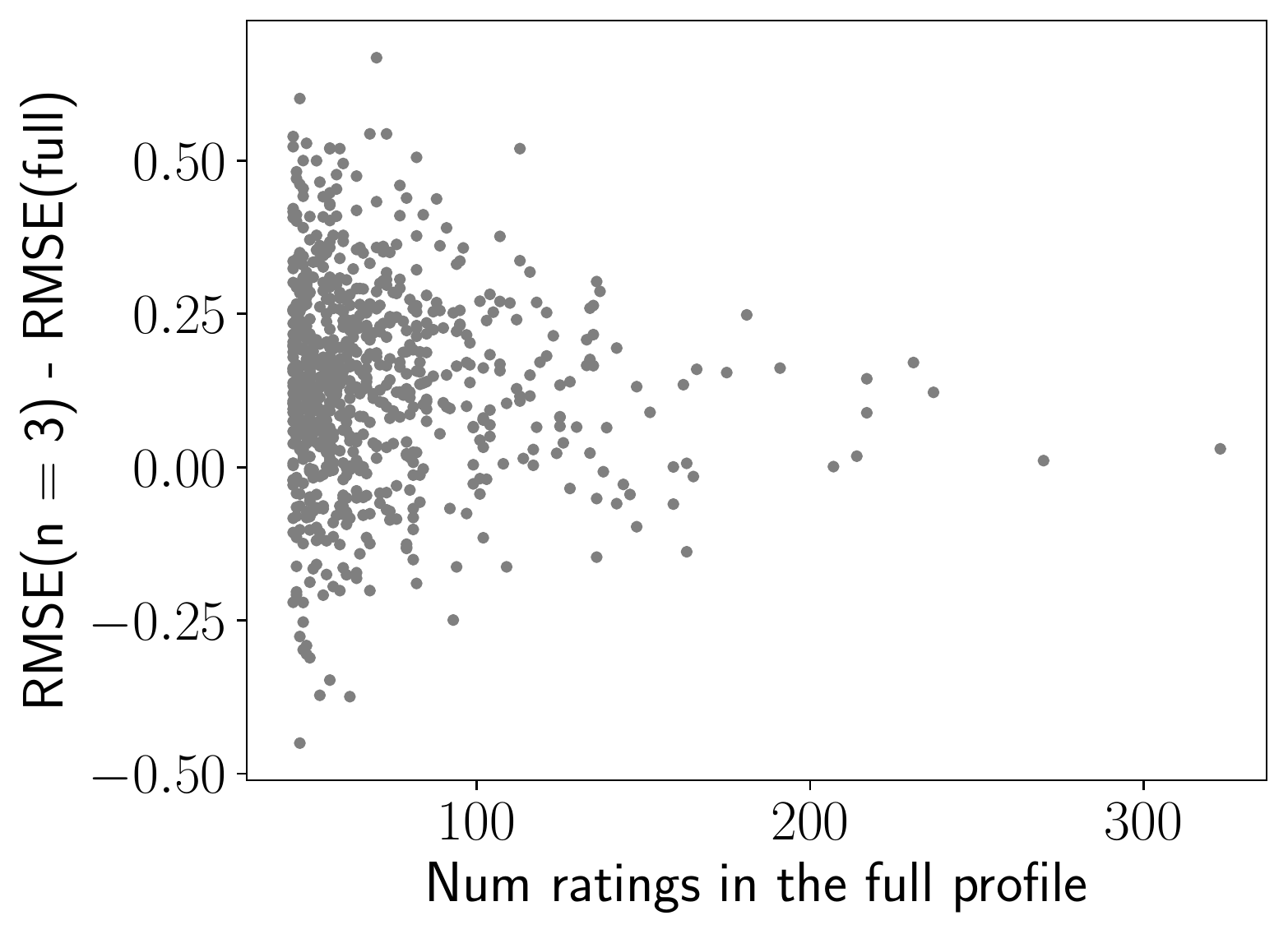}
             \caption{}
             \label{fig:num-ratings-error-delta}
     \end{subfigure}
     \hfill
      \begin{subfigure}[t]{0.2\textwidth}
             \includegraphics[width=0.98\textwidth]{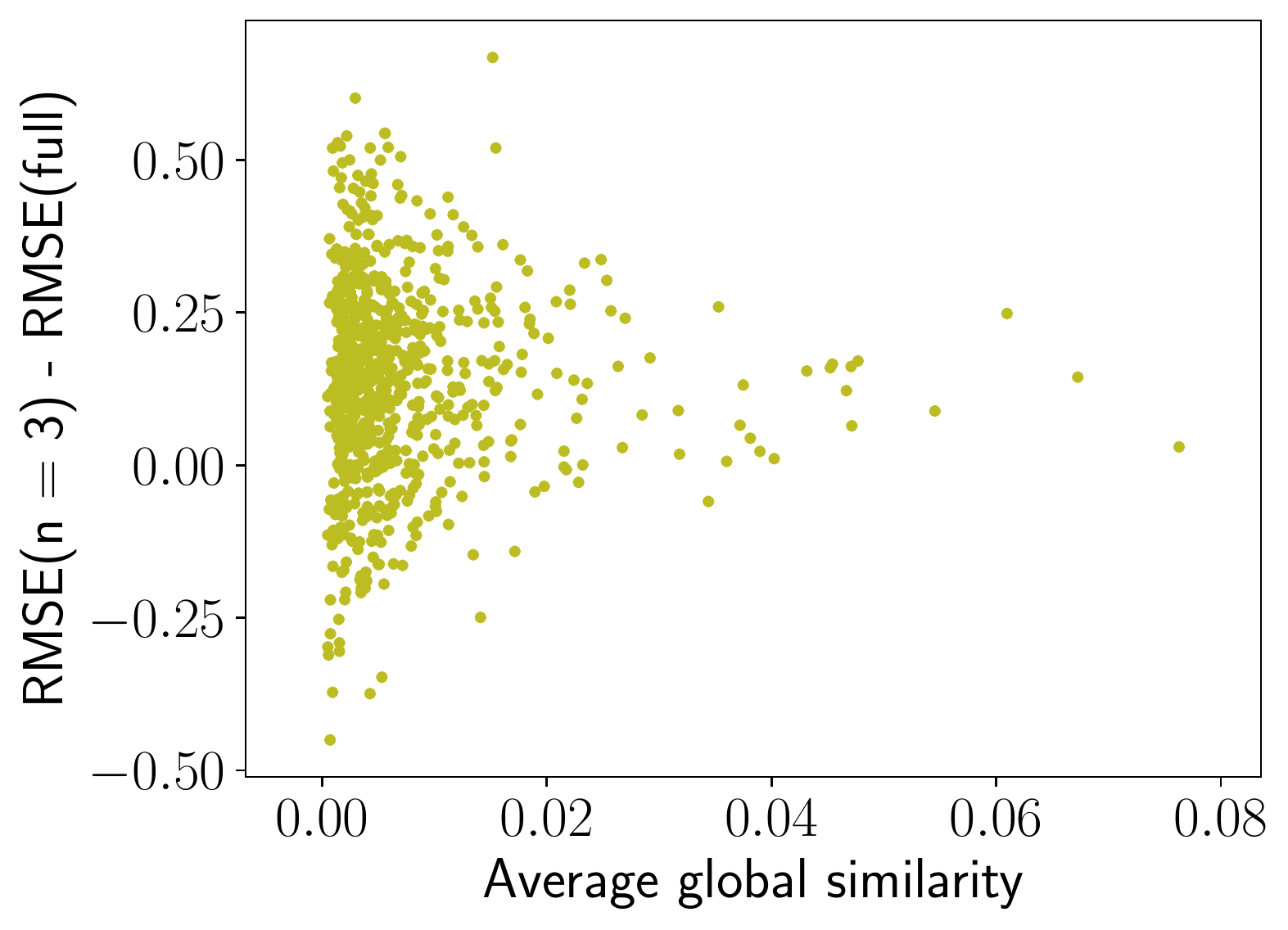}
             \caption{}
             \label{fig:avg-sim-global-error-delta}
     \end{subfigure}
     \hfill
     \begin{subfigure}[m]{0.2\textwidth}
             \includegraphics[width=0.98\textwidth]{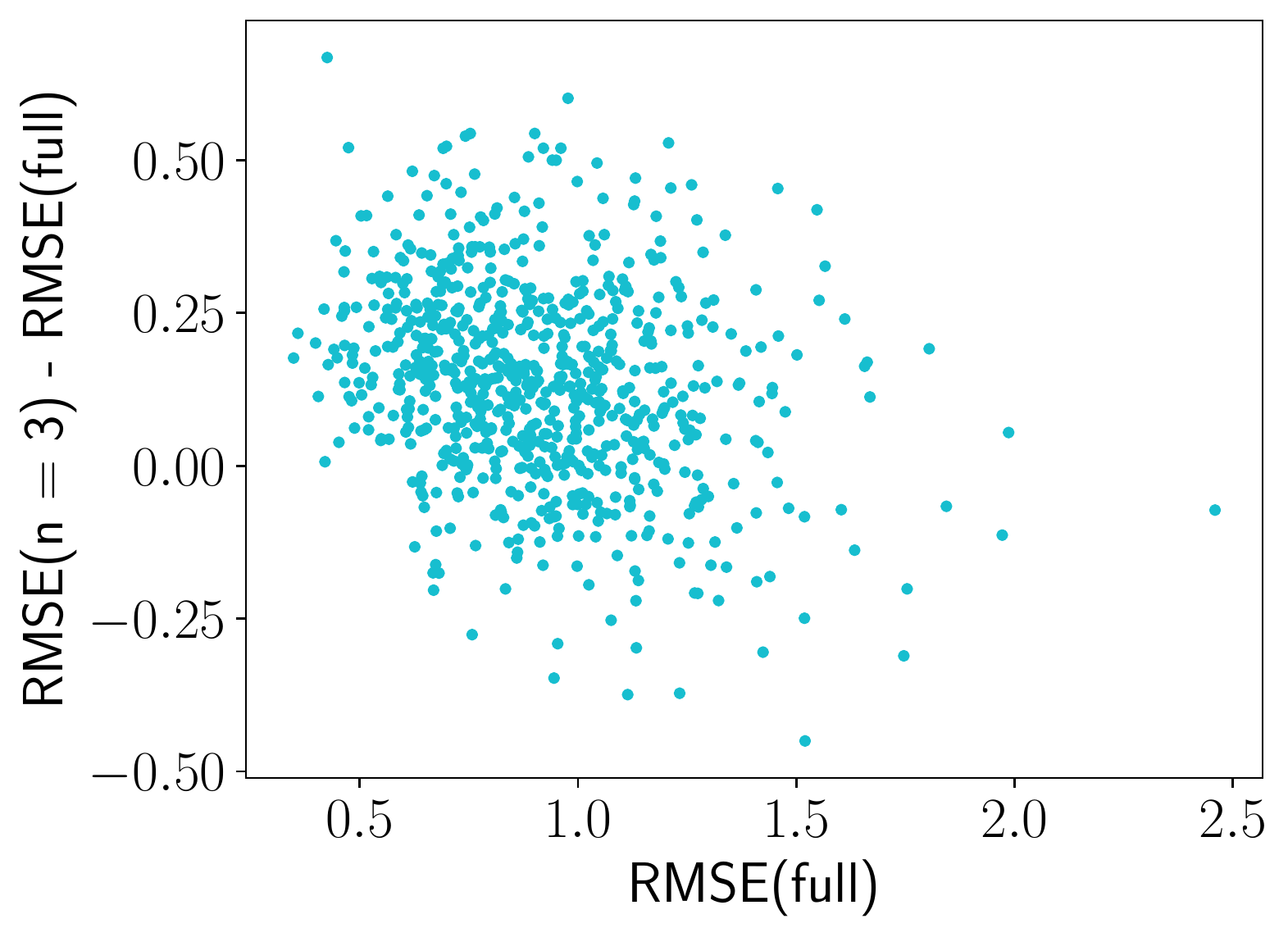}
             \caption{}
             \label{fig:rmse-full-error-delta}
     \end{subfigure}
     \hfill
     \begin{subfigure}[m]{0.2\textwidth}
             \includegraphics[width=0.98\textwidth]{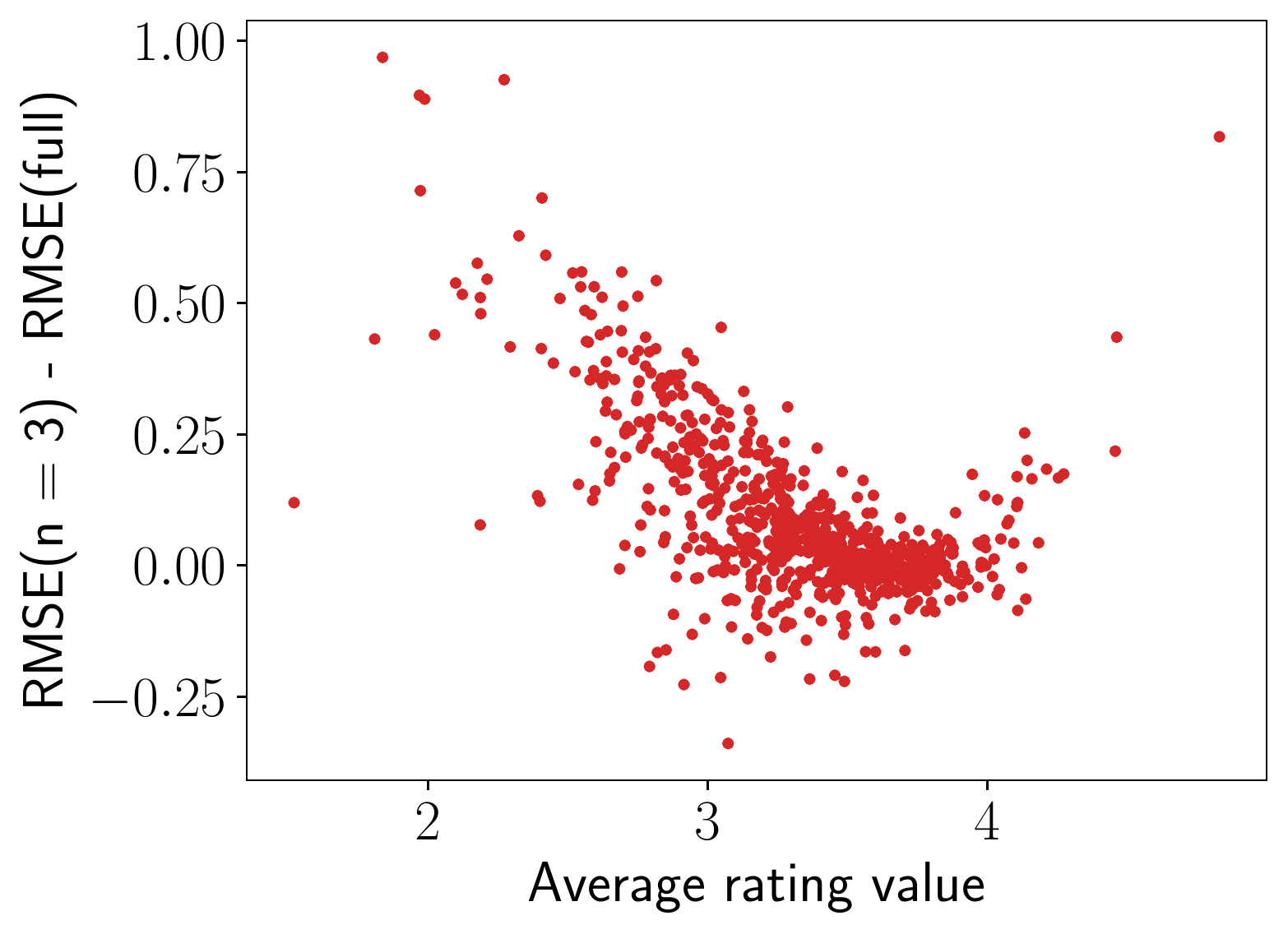}
             \caption{}
             \label{fig:avg-rating-most-error-delta}
     \end{subfigure}
     \hfill
     \begin{subfigure}[b]{0.2\textwidth}
             \includegraphics[width=0.98\textwidth]{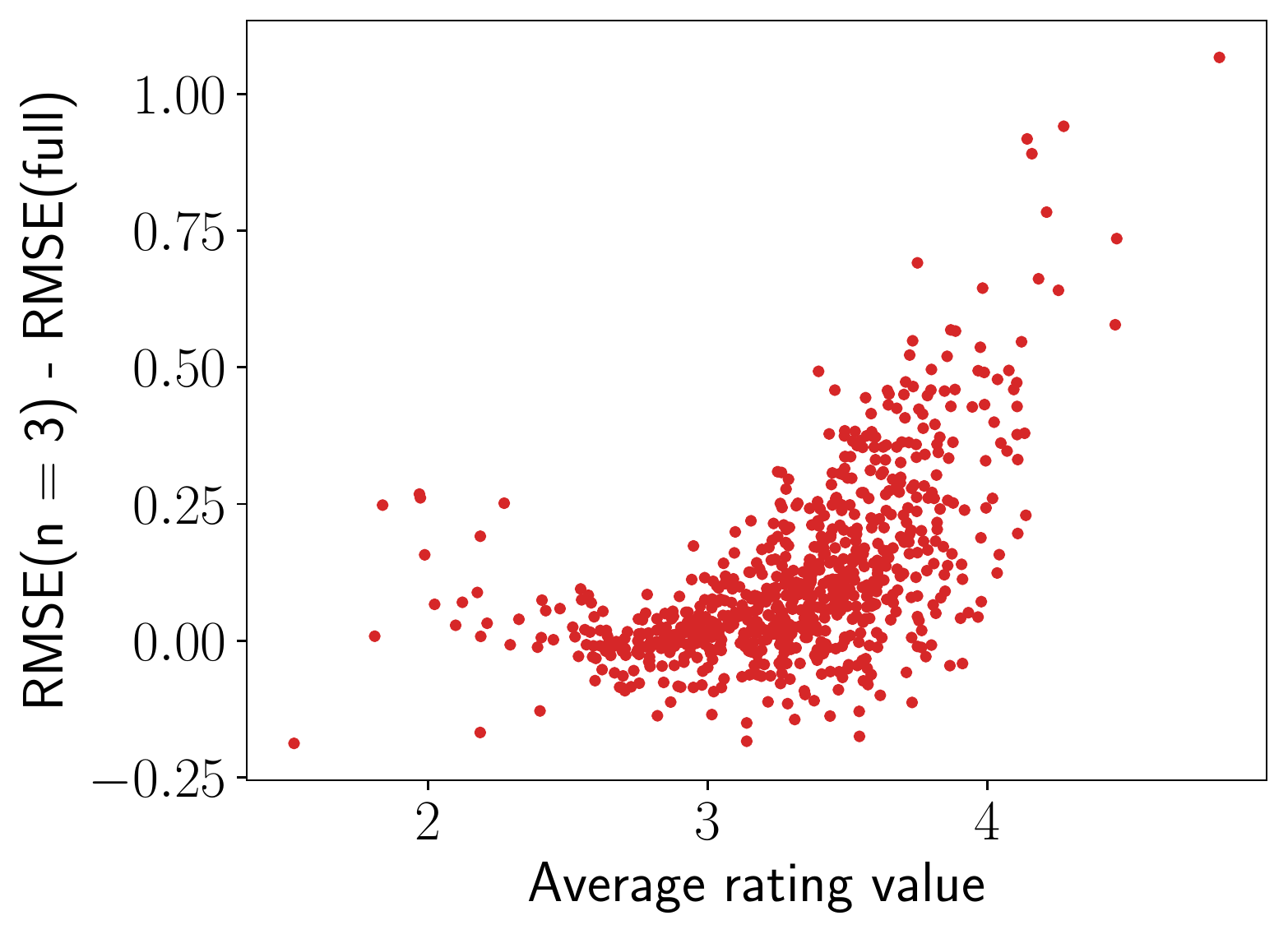}
             \caption{}
             \label{fig:avg-rating-least-error-delta}
     \end{subfigure}
     \hfill
     \begin{subfigure}[b]{0.2\textwidth}
             \includegraphics[width=0.98\textwidth]{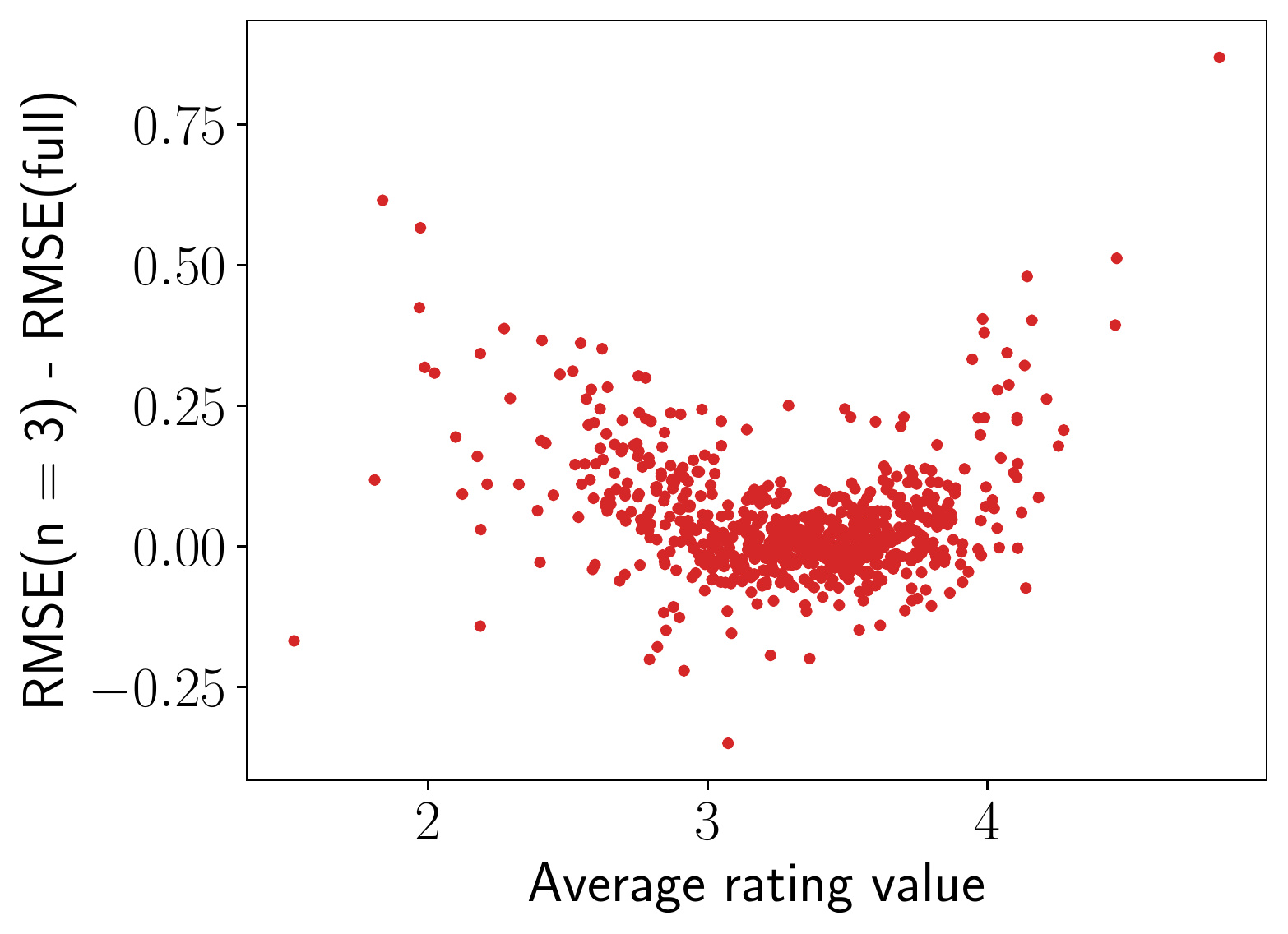}
             \caption{}
             \label{fig:avg-rating-random-error-delta}
     \end{subfigure}
\caption{Selected cases of the dependencies between user characteristics and the RMSE delta (higher value means higher quality loss for minimization) incurred by minimization: (a) Number of ratings in the full profile, kNN, Random, (b) Average similarity to all users in the data, kNN, Random, (c) RMSE of recommendations over full profile, kNN, Random (d) Average rating value in the full profile, SVD, Most Favorite, (e) Average rating value in the full profile, SVD, Least Favorite, (e) Average rating value in the full profile, SVD, Random. In each of the plots, a dot corresponds to one minimizing user.}
\label{fig:user-characteristics-vs-loss}
\end{figure}

For a closer look into the complex dependencies between user characteristics, minimization strategies, recommendation algorithms, and minimization errors, we plot the most interesting cases in Figure~\ref{fig:user-characteristics-vs-loss}. 

Figure~\ref{fig:num-ratings-error-delta} shows the dependency between the number of ratings in a user's full profile and the error delta (kNN+Random). The plot suggests that the smaller a user's observational pool, the higher variation in the incurred minimization error. We conjecture that the reason for this effect is that sparse profiles with little data are likely to misrepresent true user tastes.  

Figure~\ref{fig:avg-sim-global-error-delta} shows the dependency between a user's average similarity to all users in the system data and the error delta (kNN+Random). We observe a similar trend -- lower global similarity means higher variance in minimization error. However, the reason for this effect is likely different. Users who are similar to many system users are likely to end up in a neighborhood with accurate recommendations irrespective of which items are minimized out of their profiles.

Figure~\ref{fig:rmse-full-error-delta} shows the dependency between a user's RMSE error for recommendations over the full observational pool the error delta (kNN+Random). We observe that lower RMSE values over the full data tend to imply higher error deltas, suggesting that users who are underserved by a system will be harmed the most when minimizing data.

Figures~\ref{fig:avg-rating-most-error-delta} and~\ref{fig:avg-rating-least-error-delta} reveal a curious observation about the dependency between the average value of ratings in a user profile and the error delta incurred by the Most and Least Favorite strategies for SVD. Users who tend to give lower movie ratings on average will receive worse results when minimizing using the Most Favorite strategy -- likely because the movies they like the most will look like neutral movies when compared to the absolute values of ratings of other users. For a similar reason, though inverted, users who tend to give higher ratings on average will receive worse results when minimizing using the Least Favorite strategy. Figure~\ref{fig:avg-rating-random-error-delta} shows that for the Random strategy the effect is symmetric and less pronounced.

\section{Data Minimization vs. Privacy}
\label{sec:privacy}

The operational definitions of data minimization proposed in this paper, as shown in the experiments, will often lead to a decrease of data collection. However, it is a feasible scenario that that each data point positively contributes to a system's performance and data collection will not be decreased as a result. Rather than thinking of minimization as another computational definition of privacy, we look at data protection more broadly. For instance, the UK Information Commissioner's Office defines data protection as \emph{`the fair and proper use of information about people'}~\cite{icoDataProtectionConcepts}.
Nevertheless, because of the potential decrease in the collected data, the proposed definitions of data minimization are related to different computational concepts of privacy. We briefly discuss some of these relationships.

\noindent \textbf{Identifiablity.}
Presence of a unique combination of items in an anonymous user profile poses a deanonymizaton risk: if an attacker has the background knowledge that a user has rated these items, they can uniquely identify their profile and thus gain access to the rest of the user's data. Analogous scenarios motivated the work on \emph{k-anonymity} and related concepts~\cite{sweeney2002k}. One way of quantifying identifiability without access to external datasets is through a lower bound on the number of items an attacker would need to know to identify a user in a given dataset. More specifically, we compute, for each user $u$, the minimum size of a subset of her ratings that does not exist in a profile of any other user: 
$
\min_{I \in \mathcal{P}(I_u)} |I| \text{  s.t.  } \forall_{v\neq u} I \nsubseteq I_v.\
$
The higher the value of the above measure, the bigger the number of items an attacker would need to know to uniquely identify a user profile, and thus the lower the identifiability risk.

Table~\ref{tab:identifiability-stats} presents the identifiability statistics for user profiles minimized using different strategies, averaged over all users. The results suggest that minimization strategies selecting items based on the characteristics of system data (Most Watched, Highest Variance, Most Characteristic) lead to lower profile identifiability than minimization methods based on an individual's preferences (Most and Least Favorite).
The most Recent strategy leads to the lowest identifiability across different values of the minimization parameter $n$. We conjecture this is because at a given time, many users rate the same new releases.

\noindent \textbf{Profiling.}
Another computational privacy concept is that of profiling---collecting detailed topical profiles of users~\cite{biega2017privacy,chen2011ups,xu2007privacy}. Should data minimization lead to decrease of collected data, it is likely that profiling risks also decrease. For instance, in our experiments, decreasing the number of movie ratings in all user's profile to a maximum of 100 already reduces the average number of different genres in a user profile from $28.2$ down to $25.1$ according to the best strategy.

\noindent \textbf{Other.} While decreasing the size of data might also help with other privacy dimensions, such as protection from inference~\cite{du2012privacy} or differential privacy~\cite{dwork2008differential} in case aggregate data is released, analysis of these dimensions is more complex and might lead to removal of different data points.

\begin{table}[h!]
\tiny
\caption{Identifiability (the minimum number of items necessary to uniquely identify a user) for user profiles minimized using different strategies, averaged over all users.} 
\label{tab:identifiability-stats}
\resizebox{0.4\textwidth}{!}{%
\begin{tabular}{ c  c  c  c  c }
\toprule
 & \textbf{n=3} & \textbf{n=7} & \textbf{n=15} & \textbf{n=100}  \\ \midrule
\textbf{random} & $2.02$ & $1.89$ & $1.76$ & $1.55$ \\
\textbf{most recent} & $1.91$ & $1.79$ & $1.71$ & $1.55$ \\
\textbf{most favorite} & $2.01$ & $1.88$ & $1.79$ & $1.55$ \\
\textbf{least favorite} & $1.92$ & $1.81$ & $1.71$ & $1.55$ \\
\textbf{most watched} & $2.28$ & $2.33$ & $2.00$ & $1.57$ \\
\textbf{most characteristic} & $1.99$ & $2.00$ & $2.00$ & $1.57$ \\
\textbf{highest variance} & $2.04$ & $2.00$ & $2.00$ & $1.57$ \\
\bottomrule
\end{tabular}
}
\end{table}

\section{Related Work}
\label{sec:rel-work}

\noindent \textbf{Interpreting GDPR principles in practice.}
The core contribution of this paper is in pointing out the gap between the current understanding of GDPR's data minimization principle and the reality of personalization systems and proposing possible adequate re-interpretations. In this context, our work is related to other efforts to translate GDPR's principles into data science practice. 
Prior work in this space has explored practical challenges behind revoking consent to data processing\cite{politou2018forgetting,utz2019informed}, and explored what \emph{the right to be forgotten}~\cite{koops2011forgetting} means in practice. Recent work proposes practical solutions for removing data points from trained machine learning models in case an individual included in the training data requests deletion~\cite{ginart2019making}. The right to explanation\cite{kaminski2019right}, requiring service providers to be able to explain algorithmic decisions and results to their users, motivated the active are of explainability and transparency. 
Another line of work analyzes changes to the online ecosystem incurred by GDPR, including the presence of consent notices~\cite{urban2019study}, or tracking scripts\cite{solomos2019clash,sorensen2019before}.

\noindent \textbf{Privacy.}
As discussed in Sec.~\ref{sec:privacy}, data minimization is related to some of the computational concepts of privacy. In the context of personalized search, many works proposed mechanisms for perturbing user search logs while preserving the search quality, including mixing and merging queries into synthetic profiles\cite{biega2017privacy,eslami2017privacy}, grouping user profiles~\cite{mac20183ps}, or splitting them~\cite{chen2011ups,xu2007privacy,zhu2010anonymizing}. Privacy has also been interpreted as a probabilistic guarantee on data retention~\cite{singla2014stochastic}.
To preserve the privacy of recommender system users, it has been proposed to prevent the collection of ratings locally if they are predicted to lead to privacy loss~\cite{guerraoui2017utility}, or to store the ratings of different users intermingled~\cite{biega2017privacy}.
Research in privacy-preserving information retrieval~\cite{yang2016privacy} moreover investigates problems related to search log anonymization~\cite{zhang2016anonymizing}, or the relation between user behavior and privacy attitudes~\cite{zimmerman2019investigating}.

\noindent \textbf{Performance of recommender systems under varying conditions.}
Analyses we perform in this paper are related to a line of work analyzing the success and failure of recommender systems under changing conditions. Ekstrand et al.~\cite{ekstrand2012recommenders} analyze data factors that cause different recommendation algorithms to fail.
Chow et al.~\cite{chow2013differential} propose techniques to estimate the contributions of different data points to the overall recommendation quality. 
Vincent et al.~\cite{vincent2019data} propose 'data strikes' as a form of collective action where users protest by withholding their data from recommendation provider. Note that, while the goal of data strikes is to limit availability of data to \emph{reduce} recommendation performance, the goal of performance-based data minimization it to limit availability of data while \emph{preserving} recommendation performance.
Wen et al.~\cite{wen2018exploring} analyzed performance decrease in a recommendation privacy scenario where users provide an algorithm with their recommendation data from the most recent N days. 

\noindent \textbf{Relation to other disciplines}
We believe that further work on data minimization would lead to synergies not only with the legal community, but also with other computer science subdisciplines. While the focus of data minimization is on minimizing features rather than data points, the problem is related to works studying the relationships between training examples and algorithmic performance. This abstract description includes, for instance, the problems of data influence~\cite{koh2017understanding}, data valuation~\cite{ghorbani2019data}, active learning~\cite{bachman2017learning}, or budgeted learning~\cite{lizotte2002budgeted}.
\vspace{-5px}
\section{Discussion and Conclusions}

\subsection{Summary of the findings}
In this paper, we have identified a lack of a homogeneous interpretation of the GDPR's purpose limitation and data minimization principles in the domain of personalization systems. We argue that these systems do not necessarily need to collect user data, but that they do so in order to improve the quality of the results. Thus, we propose two performance-based interpretations of the data minimization principle that tie the limitations of data collection to quality metrics. The first interpretation focuses on the global average algorithm performance, while the second focuses on the local per-user minimum performance.

We found SVD (FunkSVD with user and item biases) to be more robust to minimization than kNN user-user collaborative filtering across different minimization strategies. Among the minimization strategies, we found the random strategy to perform well, likely due to the fact that it preserves average user characteristics. 
However, for each recommendation algorithm, it is possible to find strategies that perform better or worse than random. 

While the results suggest global data minimization can be quite successful (in some cases we can withhold as much as 90\% of the user data incurring RMSE loss as low as 0.025), we show that quality difference can be substantial for individual users. Furthermore, our analysis with Ordinal Least Squares regression shows that the variation in individual-level error is not well explained by standard user features. The complex interaction between the individual-level error and recommendation algorithms, minimization strategies, system data, and individual data, require further study, also from a legal perspective. Indeed, further research should evaluate the desirability of both approaches, considering that, on the one hand, the GDPR requires that each data processing operation be examined on its own merits, yet on the other purpose limitation or data protection by design and by default ought to be evaluated from a global perspective. 

\subsection{Potential negative impacts}
Based on our observations about varying user-level errors, it is plausible that data minimization hurts marginalized groups, in particular if those groups form a minority of the data---the members of majority population will be well served with just a few features (because there is sufficient statistical support), while minority populations will \emph{need} to provide more features to get service of comparable quality.
A scenario like this would further harm marginalized populations through decreased data protection.

Furthermore, our analysis assumes service providers have a collection of background data to base personalization on (purchased or collected from markets that are not legally obliged to data minimization). Companies might also need personal data to develop new services. In this work, we did not consider such provider costs.

\subsection{Challenges for data minimization}
While this paper enhances our understanding of what performance-based data minimization means in practice, a number of challenges emerge. Practical minimization mechanisms would not be able to measure quality loss directly, nor easily adapt selection mechanisms to each user if necessary without access to candidate user data. To support minimization, we need to design new protocols for user-system interaction, and new learning mechanisms that select data while respecting specific minimization requirements. Last but not least, further interdisciplinary work with the legal community is necessary to develop data minimization interpretations that are verifiable and viable, both legally and computationally.
\label{sec:conclusions}

\begin{acks}
We wish to thank Solon Barocas for the discussion that helped shape the interdisciplinary direction of this work.
\end{acks}

{
\bibliographystyle{ACM-Reference-Format}
\bibliography{main}


\begin{thebibliography}{48}


\ifx \showCODEN    \undefined \def \showCODEN     #1{\unskip}     \fi
\ifx \showDOI      \undefined \def \showDOI       #1{#1}\fi
\ifx \showISBNx    \undefined \def \showISBNx     #1{\unskip}     \fi
\ifx \showISBNxiii \undefined \def \showISBNxiii  #1{\unskip}     \fi
\ifx \showISSN     \undefined \def \showISSN      #1{\unskip}     \fi
\ifx \showLCCN     \undefined \def \showLCCN      #1{\unskip}     \fi
\ifx \shownote     \undefined \def \shownote      #1{#1}          \fi
\ifx \showarticletitle \undefined \def \showarticletitle #1{#1}   \fi
\ifx \showURL      \undefined \def \showURL       {\relax}        \fi
\providecommand\bibfield[2]{#2}
\providecommand\bibinfo[2]{#2}
\providecommand\natexlab[1]{#1}
\providecommand\showeprint[2][]{arXiv:#2}

\bibitem[\protect\citeauthoryear{Bachman, Sordoni, and Trischler}{Bachman
  et~al\mbox{.}}{2017}]%
        {bachman2017learning}
\bibfield{author}{\bibinfo{person}{Philip Bachman}, \bibinfo{person}{Alessandro
  Sordoni}, {and} \bibinfo{person}{Adam Trischler}.}
  \bibinfo{year}{2017}\natexlab{}.
\newblock \showarticletitle{Learning algorithms for active learning}. In
  \bibinfo{booktitle}{\emph{Proceedings of the 34th International Conference on
  Machine Learning-Volume 70}}. \bibinfo{pages}{301--310}.
\newblock


\bibitem[\protect\citeauthoryear{Biega, Saha~Roy, and Weikum}{Biega
  et~al\mbox{.}}{2017}]%
        {biega2017privacy}
\bibfield{author}{\bibinfo{person}{Asia~J Biega}, \bibinfo{person}{Rishiraj
  Saha~Roy}, {and} \bibinfo{person}{Gerhard Weikum}.}
  \bibinfo{year}{2017}\natexlab{}.
\newblock \showarticletitle{Privacy through solidarity: A
  user-utility-preserving framework to counter profiling}. In
  \bibinfo{booktitle}{\emph{Proceedings of the 40th International ACM SIGIR
  Conference on Research and Development in Information Retrieval}}. ACM,
  \bibinfo{pages}{675--684}.
\newblock


\bibitem[\protect\citeauthoryear{Binns and Gallo}{Binns and Gallo}{2019}]%
        {icoDataMinimizationTechniques}
\bibfield{author}{\bibinfo{person}{Reuben Binns} {and} \bibinfo{person}{Valeria
  Gallo}.} \bibinfo{year}{2019}\natexlab{}.
\newblock \bibinfo{booktitle}{\emph{Data minimisation and privacy-preserving
  techniques in AI systems}}.
\newblock
\urldef\tempurl%
\url{https://ico.org.uk/about-the-ico/news-and-events/ai-blog-data-minimisation-and-privacy-preserving-techniques-in-ai-systems/}
\showURL{%
Retrieved May 26, 2020 from \tempurl}


\bibitem[\protect\citeauthoryear{Chen, Bai, Shou, Chen, and Gao}{Chen
  et~al\mbox{.}}{2011}]%
        {chen2011ups}
\bibfield{author}{\bibinfo{person}{Gang Chen}, \bibinfo{person}{He Bai},
  \bibinfo{person}{Lidan Shou}, \bibinfo{person}{Ke Chen}, {and}
  \bibinfo{person}{Yunjun Gao}.} \bibinfo{year}{2011}\natexlab{}.
\newblock \showarticletitle{UPS: efficient privacy protection in personalized
  web search}. In \bibinfo{booktitle}{\emph{Proceedings of the 34th
  international ACM SIGIR conference on Research and development in Information
  Retrieval}}. ACM, \bibinfo{pages}{615--624}.
\newblock


\bibitem[\protect\citeauthoryear{Chow, Jin, Knijnenburg, and Saldamli}{Chow
  et~al\mbox{.}}{2013}]%
        {chow2013differential}
\bibfield{author}{\bibinfo{person}{Richard Chow}, \bibinfo{person}{Hongxia
  Jin}, \bibinfo{person}{Bart Knijnenburg}, {and} \bibinfo{person}{Gokay
  Saldamli}.} \bibinfo{year}{2013}\natexlab{}.
\newblock \showarticletitle{Differential data analysis for recommender
  systems}. In \bibinfo{booktitle}{\emph{Proceedings of the 7th ACM conference
  on Recommender systems}}. \bibinfo{pages}{323--326}.
\newblock


\bibitem[\protect\citeauthoryear{du~Pin~Calmon and Fawaz}{du~Pin~Calmon and
  Fawaz}{2012}]%
        {du2012privacy}
\bibfield{author}{\bibinfo{person}{Fl{\'a}vio du Pin~Calmon} {and}
  \bibinfo{person}{Nadia Fawaz}.} \bibinfo{year}{2012}\natexlab{}.
\newblock \showarticletitle{Privacy against statistical inference}. In
  \bibinfo{booktitle}{\emph{2012 50th annual Allerton conference on
  communication, control, and computing (Allerton)}}. IEEE,
  \bibinfo{pages}{1401--1408}.
\newblock


\bibitem[\protect\citeauthoryear{Dwork}{Dwork}{2008}]%
        {dwork2008differential}
\bibfield{author}{\bibinfo{person}{Cynthia Dwork}.}
  \bibinfo{year}{2008}\natexlab{}.
\newblock \showarticletitle{Differential privacy: A survey of results}. In
  \bibinfo{booktitle}{\emph{International conference on theory and applications
  of models of computation}}. Springer, \bibinfo{pages}{1--19}.
\newblock


\bibitem[\protect\citeauthoryear{Ekstrand and Riedl}{Ekstrand and
  Riedl}{2012}]%
        {ekstrand2012recommenders}
\bibfield{author}{\bibinfo{person}{Michael Ekstrand} {and}
  \bibinfo{person}{John Riedl}.} \bibinfo{year}{2012}\natexlab{}.
\newblock \showarticletitle{When recommenders fail: predicting recommender
  failure for algorithm selection and combination}. In
  \bibinfo{booktitle}{\emph{Proceedings of the sixth ACM conference on
  Recommender systems}}. ACM, \bibinfo{pages}{233--236}.
\newblock


\bibitem[\protect\citeauthoryear{Eslami, Biega, Saha~Roy, and Weikum}{Eslami
  et~al\mbox{.}}{2017}]%
        {eslami2017privacy}
\bibfield{author}{\bibinfo{person}{Sedigheh Eslami}, \bibinfo{person}{Asia~J
  Biega}, \bibinfo{person}{Rishiraj Saha~Roy}, {and} \bibinfo{person}{Gerhard
  Weikum}.} \bibinfo{year}{2017}\natexlab{}.
\newblock \showarticletitle{Privacy of hidden profiles: Utility-preserving
  profile removal in online forums}. In \bibinfo{booktitle}{\emph{Proceedings
  of the 2017 ACM on Conference on Information and Knowledge Management}}. ACM,
  \bibinfo{pages}{2063--2066}.
\newblock


\bibitem[\protect\citeauthoryear{Finck and Pallas}{Finck and Pallas}{2019}]%
        {finck2019they}
\bibfield{author}{\bibinfo{person}{Mich{\`e}le Finck} {and}
  \bibinfo{person}{Frank Pallas}.} \bibinfo{year}{2019}\natexlab{}.
\newblock \showarticletitle{They Who Must Not Be Identified-Distinguishing
  Personal from Non-Personal Data Under the GDPR}.
\newblock \bibinfo{journal}{\emph{Max Planck Institute for Innovation \&
  Competition Research Paper}} \bibinfo{number}{19-14} (\bibinfo{year}{2019}).
\newblock


\bibitem[\protect\citeauthoryear{Frolov and Oseledets}{Frolov and
  Oseledets}{2016}]%
        {frolov2016fifty}
\bibfield{author}{\bibinfo{person}{Evgeny Frolov} {and} \bibinfo{person}{Ivan
  Oseledets}.} \bibinfo{year}{2016}\natexlab{}.
\newblock \showarticletitle{Fifty Shades of Ratings: How to Benefit from a
  Negative Feedback in Top-N Recommendations Tasks}. In
  \bibinfo{booktitle}{\emph{Proceedings of the 10th ACM Conference on
  Recommender Systems}}. ACM, \bibinfo{pages}{91--98}.
\newblock


\bibitem[\protect\citeauthoryear{Funk}{Funk}{2006}]%
        {funk2006netflix}
\bibfield{author}{\bibinfo{person}{Simon Funk}.}
  \bibinfo{year}{2006}\natexlab{}.
\newblock \showarticletitle{Netflix update: Try this at home (December 2006)}.
\newblock \bibinfo{journal}{\emph{URL http://sifter. org/\~{}
  simon/journal/20061211. html}} (\bibinfo{year}{2006}).
\newblock


\bibitem[\protect\citeauthoryear{Galdon~Clavell, Mart{\'\i}n~Zamorano,
  Castillo, Smith, and Matic}{Galdon~Clavell et~al\mbox{.}}{2020}]%
        {galdon2020auditing}
\bibfield{author}{\bibinfo{person}{Gemma Galdon~Clavell},
  \bibinfo{person}{Mariano Mart{\'\i}n~Zamorano}, \bibinfo{person}{Carlos
  Castillo}, \bibinfo{person}{Oliver Smith}, {and} \bibinfo{person}{Aleksandar
  Matic}.} \bibinfo{year}{2020}\natexlab{}.
\newblock \showarticletitle{Auditing Algorithms: On Lessons Learned and the
  Risks of Data Minimization}. In \bibinfo{booktitle}{\emph{Proceedings of the
  AAAI/ACM Conference on AI, Ethics, and Society}}. \bibinfo{pages}{265--271}.
\newblock


\bibitem[\protect\citeauthoryear{Ghorbani and Zou}{Ghorbani and Zou}{2019}]%
        {ghorbani2019data}
\bibfield{author}{\bibinfo{person}{Amirata Ghorbani} {and}
  \bibinfo{person}{James Zou}.} \bibinfo{year}{2019}\natexlab{}.
\newblock \showarticletitle{Data Shapley: Equitable Valuation of Data for
  Machine Learning}. In \bibinfo{booktitle}{\emph{International Conference on
  Machine Learning}}. \bibinfo{pages}{2242--2251}.
\newblock


\bibitem[\protect\citeauthoryear{Ginart, Guan, Valiant, and Zou}{Ginart
  et~al\mbox{.}}{2019}]%
        {ginart2019making}
\bibfield{author}{\bibinfo{person}{Antonio Ginart}, \bibinfo{person}{Melody
  Guan}, \bibinfo{person}{Gregory Valiant}, {and} \bibinfo{person}{James Zou}.}
  \bibinfo{year}{2019}\natexlab{}.
\newblock \showarticletitle{Making AI Forget You: Data Deletion in Machine
  Learning}.
\newblock \bibinfo{journal}{\emph{NIPS}} (\bibinfo{year}{2019}).
\newblock


\bibitem[\protect\citeauthoryear{Guerraoui, Kermarrec, and Taziki}{Guerraoui
  et~al\mbox{.}}{2017}]%
        {guerraoui2017utility}
\bibfield{author}{\bibinfo{person}{Rachid Guerraoui},
  \bibinfo{person}{Anne-Marie Kermarrec}, {and} \bibinfo{person}{Mahsa
  Taziki}.} \bibinfo{year}{2017}\natexlab{}.
\newblock \showarticletitle{The Utility and Privacy Effects of a Click}. In
  \bibinfo{booktitle}{\emph{Proceedings of the 40th International ACM SIGIR
  Conference on Research and Development in Information Retrieval}}. ACM,
  \bibinfo{pages}{665--674}.
\newblock


\bibitem[\protect\citeauthoryear{Guyon and Elisseeff}{Guyon and
  Elisseeff}{2003}]%
        {guyon2003introduction}
\bibfield{author}{\bibinfo{person}{Isabelle Guyon} {and}
  \bibinfo{person}{Andr{\'e} Elisseeff}.} \bibinfo{year}{2003}\natexlab{}.
\newblock \showarticletitle{An introduction to variable and feature selection}.
\newblock \bibinfo{journal}{\emph{Journal of machine learning research}}
  \bibinfo{volume}{3}, \bibinfo{number}{Mar} (\bibinfo{year}{2003}),
  \bibinfo{pages}{1157--1182}.
\newblock


\bibitem[\protect\citeauthoryear{{Halevy}, {Norvig}, and {Pereira}}{{Halevy}
  et~al\mbox{.}}{2009}]%
        {halevy:unreasonable-data}
\bibfield{author}{\bibinfo{person}{A. {Halevy}}, \bibinfo{person}{P. {Norvig}},
  {and} \bibinfo{person}{F. {Pereira}}.} \bibinfo{year}{2009}\natexlab{}.
\newblock \showarticletitle{The Unreasonable Effectiveness of Data}.
\newblock \bibinfo{journal}{\emph{IEEE Intelligent Systems}}
  \bibinfo{volume}{24}, \bibinfo{number}{2} (\bibinfo{date}{March}
  \bibinfo{year}{2009}), \bibinfo{pages}{8--12}.
\newblock


\bibitem[\protect\citeauthoryear{Harper and Konstan}{Harper and
  Konstan}{2016}]%
        {harper2016movielens}
\bibfield{author}{\bibinfo{person}{F~Maxwell Harper} {and}
  \bibinfo{person}{Joseph~A Konstan}.} \bibinfo{year}{2016}\natexlab{}.
\newblock \showarticletitle{The movielens datasets: History and context}.
\newblock \bibinfo{journal}{\emph{Acm transactions on interactive intelligent
  systems (tiis)}} \bibinfo{volume}{5}, \bibinfo{number}{4}
  (\bibinfo{year}{2016}), \bibinfo{pages}{19}.
\newblock


\bibitem[\protect\citeauthoryear{He, Kang, and McAuley}{He
  et~al\mbox{.}}{2017}]%
        {he2017translation}
\bibfield{author}{\bibinfo{person}{Ruining He}, \bibinfo{person}{Wang-Cheng
  Kang}, {and} \bibinfo{person}{Julian McAuley}.}
  \bibinfo{year}{2017}\natexlab{}.
\newblock \showarticletitle{Translation-based recommendation}. In
  \bibinfo{booktitle}{\emph{Proceedings of the Eleventh ACM Conference on
  Recommender Systems}}. \bibinfo{pages}{161--169}.
\newblock


\bibitem[\protect\citeauthoryear{Hug}{Hug}{2017}]%
        {Surprise}
\bibfield{author}{\bibinfo{person}{Nicolas Hug}.}
  \bibinfo{year}{2017}\natexlab{}.
\newblock \bibinfo{title}{{S}urprise, a {P}ython library for recommender
  systems}.
\newblock \bibinfo{howpublished}{\url{http://surpriselib.com}}.
\newblock


\bibitem[\protect\citeauthoryear{J\"{a}rvelin and
  Kek\"{a}l\"{a}inen}{J\"{a}rvelin and Kek\"{a}l\"{a}inen}{2002}]%
        {NDCG}
\bibfield{author}{\bibinfo{person}{Kalervo J\"{a}rvelin} {and}
  \bibinfo{person}{Jaana Kek\"{a}l\"{a}inen}.} \bibinfo{year}{2002}\natexlab{}.
\newblock \showarticletitle{Cumulated gain-based evaluation of IR techniques}.
\newblock \bibinfo{journal}{\emph{TOIS}} \bibinfo{volume}{20},
  \bibinfo{number}{4} (\bibinfo{year}{2002}), \bibinfo{pages}{422--446}.
\newblock


\bibitem[\protect\citeauthoryear{Kaminski}{Kaminski}{2019}]%
        {kaminski2019right}
\bibfield{author}{\bibinfo{person}{Margot~E Kaminski}.}
  \bibinfo{year}{2019}\natexlab{}.
\newblock \showarticletitle{The right to explanation, explained}.
\newblock \bibinfo{journal}{\emph{Berkeley Tech. LJ}}  \bibinfo{volume}{34}
  (\bibinfo{year}{2019}), \bibinfo{pages}{189}.
\newblock


\bibitem[\protect\citeauthoryear{Koh and Liang}{Koh and Liang}{2017}]%
        {koh2017understanding}
\bibfield{author}{\bibinfo{person}{Pang~Wei Koh} {and} \bibinfo{person}{Percy
  Liang}.} \bibinfo{year}{2017}\natexlab{}.
\newblock \showarticletitle{Understanding black-box predictions via influence
  functions}. In \bibinfo{booktitle}{\emph{Proceedings of the 34th
  International Conference on Machine Learning-Volume 70}}.
  \bibinfo{pages}{1885--1894}.
\newblock


\bibitem[\protect\citeauthoryear{Koops}{Koops}{2011}]%
        {koops2011forgetting}
\bibfield{author}{\bibinfo{person}{Bert-Jaap Koops}.}
  \bibinfo{year}{2011}\natexlab{}.
\newblock \showarticletitle{Forgetting footprints, shunning shadows: A critical
  analysis of the right to be forgotten in big data practice}.
\newblock \bibinfo{journal}{\emph{SCRIPTed}}  \bibinfo{volume}{8}
  (\bibinfo{year}{2011}), \bibinfo{pages}{229}.
\newblock


\bibitem[\protect\citeauthoryear{Krause and Horvitz}{Krause and
  Horvitz}{2010}]%
        {krause2010utility}
\bibfield{author}{\bibinfo{person}{Andreas Krause} {and} \bibinfo{person}{Eric
  Horvitz}.} \bibinfo{year}{2010}\natexlab{}.
\newblock \showarticletitle{A utility-theoretic approach to privacy in online
  services}.
\newblock \bibinfo{journal}{\emph{Journal of Artificial Intelligence Research}}
   \bibinfo{volume}{39} (\bibinfo{year}{2010}), \bibinfo{pages}{633--662}.
\newblock


\bibitem[\protect\citeauthoryear{Lizotte, Madani, and Greiner}{Lizotte
  et~al\mbox{.}}{2002}]%
        {lizotte2002budgeted}
\bibfield{author}{\bibinfo{person}{Daniel~J Lizotte}, \bibinfo{person}{Omid
  Madani}, {and} \bibinfo{person}{Russell Greiner}.}
  \bibinfo{year}{2002}\natexlab{}.
\newblock \showarticletitle{Budgeted learning of nailve-bayes classifiers}. In
  \bibinfo{booktitle}{\emph{Proceedings of the Nineteenth conference on
  Uncertainty in Artificial Intelligence}}. \bibinfo{pages}{378--385}.
\newblock


\bibitem[\protect\citeauthoryear{Mac~Aonghusa and Leith}{Mac~Aonghusa and
  Leith}{2018}]%
        {mac20183ps}
\bibfield{author}{\bibinfo{person}{Pol Mac~Aonghusa} {and}
  \bibinfo{person}{Douglas Leith}.} \bibinfo{year}{2018}\natexlab{}.
\newblock \showarticletitle{3PS-Online Privacy through Group Identities}.
\newblock \bibinfo{journal}{\emph{arXiv preprint arXiv:1811.11039}}
  (\bibinfo{year}{2018}).
\newblock


\bibitem[\protect\citeauthoryear{Narayanan and Shmatikov}{Narayanan and
  Shmatikov}{2008}]%
        {narayanan2008robust}
\bibfield{author}{\bibinfo{person}{Arvind Narayanan} {and}
  \bibinfo{person}{Vitaly Shmatikov}.} \bibinfo{year}{2008}\natexlab{}.
\newblock \showarticletitle{Robust de-anonymization of large sparse datasets}.
  In \bibinfo{booktitle}{\emph{2008 IEEE Symposium on Security and Privacy,
  SP}}. \bibinfo{pages}{111--125}.
\newblock


\bibitem[\protect\citeauthoryear{Office}{Office}{2018a}]%
        {icoDataProtectionConcepts}
\bibfield{author}{\bibinfo{person}{UK~Information~Commisioner's Office}.}
  \bibinfo{year}{2018}\natexlab{a}.
\newblock \bibinfo{booktitle}{\emph{Guide to Data Protection. Some basic
  concepts}}.
\newblock
\urldef\tempurl%
\url{https://ico.org.uk/for-organisations/guide-to-data-protection/introduction-to-data-protection/some-basic-concepts/}
\showURL{%
Retrieved Jan 22, 2020 from \tempurl}


\bibitem[\protect\citeauthoryear{Office}{Office}{2018b}]%
        {icoDataMinimization}
\bibfield{author}{\bibinfo{person}{UK~Information~Commisioner's Office}.}
  \bibinfo{year}{2018}\natexlab{b}.
\newblock \bibinfo{booktitle}{\emph{Guide to the General Data Protection
  Regulation (GDPR). Principle (c): Data minimisation.}}
\newblock
\urldef\tempurl%
\url{https://ico.org.uk/for-organisations/guide-to-data-protection/guide-to-the-general-data-protection-regulation-gdpr/principles/data-minimisation/}
\showURL{%
Retrieved Jan 22, 2020 from \tempurl}


\bibitem[\protect\citeauthoryear{Politou, Alepis, and Patsakis}{Politou
  et~al\mbox{.}}{2018}]%
        {politou2018forgetting}
\bibfield{author}{\bibinfo{person}{Eugenia Politou}, \bibinfo{person}{Efthimios
  Alepis}, {and} \bibinfo{person}{Constantinos Patsakis}.}
  \bibinfo{year}{2018}\natexlab{}.
\newblock \showarticletitle{Forgetting personal data and revoking consent under
  the GDPR: Challenges and proposed solutions}.
\newblock \bibinfo{journal}{\emph{Journal of Cybersecurity}}
  \bibinfo{volume}{4}, \bibinfo{number}{1} (\bibinfo{year}{2018}),
  \bibinfo{pages}{tyy001}.
\newblock


\bibitem[\protect\citeauthoryear{Regulation}{Regulation}{2016}]%
        {regulation2016regulation}
\bibfield{author}{\bibinfo{person}{Protection Regulation}.}
  \bibinfo{year}{2016}\natexlab{}.
\newblock \showarticletitle{REGULATION (EU) 2016/679 OF THE EUROPEAN PARLIAMENT
  AND OF THE COUNCIL}.
\newblock \bibinfo{journal}{\emph{Official Journal of the European Union}}
  (\bibinfo{year}{2016}).
\newblock


\bibitem[\protect\citeauthoryear{Singla, Horvitz, Kamar, and White}{Singla
  et~al\mbox{.}}{2014}]%
        {singla2014stochastic}
\bibfield{author}{\bibinfo{person}{Adish Singla}, \bibinfo{person}{Eric
  Horvitz}, \bibinfo{person}{Ece Kamar}, {and} \bibinfo{person}{Ryen White}.}
  \bibinfo{year}{2014}\natexlab{}.
\newblock \showarticletitle{Stochastic privacy}. In
  \bibinfo{booktitle}{\emph{Twenty-Eighth AAAI Conference on Artificial
  Intelligence}}.
\newblock


\bibitem[\protect\citeauthoryear{Solomos, Ilia, Ioannidis, and
  Kourtellis}{Solomos et~al\mbox{.}}{2019}]%
        {solomos2019clash}
\bibfield{author}{\bibinfo{person}{Konstantinos Solomos},
  \bibinfo{person}{Panagiotis Ilia}, \bibinfo{person}{Sotiris Ioannidis}, {and}
  \bibinfo{person}{Nicolas Kourtellis}.} \bibinfo{year}{2019}\natexlab{}.
\newblock \showarticletitle{Clash of the Trackers: Measuring the Evolution of
  the Online Tracking Ecosystem}.
\newblock \bibinfo{journal}{\emph{arXiv preprint arXiv:1907.12860}}
  (\bibinfo{year}{2019}).
\newblock


\bibitem[\protect\citeauthoryear{S{\o}rensen and Kosta}{S{\o}rensen and
  Kosta}{2019}]%
        {sorensen2019before}
\bibfield{author}{\bibinfo{person}{Jannick S{\o}rensen} {and}
  \bibinfo{person}{Sokol Kosta}.} \bibinfo{year}{2019}\natexlab{}.
\newblock \showarticletitle{Before and After GDPR: The Changes in Third Party
  Presence at Public and Private European Websites}. In
  \bibinfo{booktitle}{\emph{The World Wide Web Conference}}. ACM,
  \bibinfo{pages}{1590--1600}.
\newblock


\bibitem[\protect\citeauthoryear{{Sun}, {Shrivastava}, {Singh}, and
  {Gupta}}{{Sun} et~al\mbox{.}}{2017}]%
        {sun:unreasonable-data}
\bibfield{author}{\bibinfo{person}{C. {Sun}}, \bibinfo{person}{A.
  {Shrivastava}}, \bibinfo{person}{S. {Singh}}, {and} \bibinfo{person}{A.
  {Gupta}}.} \bibinfo{year}{2017}\natexlab{}.
\newblock \showarticletitle{Revisiting Unreasonable Effectiveness of Data in
  Deep Learning Era}. In \bibinfo{booktitle}{\emph{2017 IEEE International
  Conference on Computer Vision (ICCV)}}. \bibinfo{pages}{843--852}.
\newblock


\bibitem[\protect\citeauthoryear{Sweeney}{Sweeney}{2002}]%
        {sweeney2002k}
\bibfield{author}{\bibinfo{person}{Latanya Sweeney}.}
  \bibinfo{year}{2002}\natexlab{}.
\newblock \showarticletitle{k-anonymity: A model for protecting privacy}.
\newblock \bibinfo{journal}{\emph{International Journal of Uncertainty,
  Fuzziness and Knowledge-Based Systems}} \bibinfo{volume}{10},
  \bibinfo{number}{05} (\bibinfo{year}{2002}), \bibinfo{pages}{557--570}.
\newblock


\bibitem[\protect\citeauthoryear{Urban, Tatang, Degeling, Holz, and
  Pohlmann}{Urban et~al\mbox{.}}{2019}]%
        {urban2019study}
\bibfield{author}{\bibinfo{person}{Tobias Urban}, \bibinfo{person}{Dennis
  Tatang}, \bibinfo{person}{Martin Degeling}, \bibinfo{person}{Thorsten Holz},
  {and} \bibinfo{person}{Norbert Pohlmann}.} \bibinfo{year}{2019}\natexlab{}.
\newblock \showarticletitle{A Study on Subject Data Access in Online
  Advertising After the GDPR}.
\newblock In \bibinfo{booktitle}{\emph{Data Privacy Management,
  Cryptocurrencies and Blockchain Technology}}. \bibinfo{publisher}{Springer},
  \bibinfo{pages}{61--79}.
\newblock


\bibitem[\protect\citeauthoryear{Utz, Degeling, Fahl, Schaub, and Holz}{Utz
  et~al\mbox{.}}{2019}]%
        {utz2019informed}
\bibfield{author}{\bibinfo{person}{Christine Utz}, \bibinfo{person}{Martin
  Degeling}, \bibinfo{person}{Sascha Fahl}, \bibinfo{person}{Florian Schaub},
  {and} \bibinfo{person}{Thorsten Holz}.} \bibinfo{year}{2019}\natexlab{}.
\newblock \showarticletitle{(Un) informed Consent: Studying GDPR Consent
  Notices in the Field}. In \bibinfo{booktitle}{\emph{ACM SIGSAC Conference on
  Computer and Communications Security (CCS '19)}}.
\newblock


\bibitem[\protect\citeauthoryear{Vapnik}{Vapnik}{1992}]%
        {vapnik1992principles}
\bibfield{author}{\bibinfo{person}{Vladimir Vapnik}.}
  \bibinfo{year}{1992}\natexlab{}.
\newblock \showarticletitle{Principles of risk minimization for learning
  theory}. In \bibinfo{booktitle}{\emph{Advances in neural information
  processing systems}}. \bibinfo{pages}{831--838}.
\newblock


\bibitem[\protect\citeauthoryear{Vincent, Hecht, and Sen}{Vincent
  et~al\mbox{.}}{2019}]%
        {vincent2019data}
\bibfield{author}{\bibinfo{person}{Nicholas Vincent}, \bibinfo{person}{Brent
  Hecht}, {and} \bibinfo{person}{Shilad Sen}.} \bibinfo{year}{2019}\natexlab{}.
\newblock \showarticletitle{``Data Strikes'': Evaluating the Effectiveness of a
  New Form of Collective Action Against Technology Companies}. In
  \bibinfo{booktitle}{\emph{The World Wide Web Conference}}. ACM,
  \bibinfo{pages}{1931--1943}.
\newblock


\bibitem[\protect\citeauthoryear{Wen, Yang, Sobolev, and Estrin}{Wen
  et~al\mbox{.}}{2018}]%
        {wen2018exploring}
\bibfield{author}{\bibinfo{person}{Hongyi Wen}, \bibinfo{person}{Longqi Yang},
  \bibinfo{person}{Michael Sobolev}, {and} \bibinfo{person}{Deborah Estrin}.}
  \bibinfo{year}{2018}\natexlab{}.
\newblock \showarticletitle{Exploring recommendations under user-controlled
  data filtering}. In \bibinfo{booktitle}{\emph{Proceedings of the 12th ACM
  Conference on Recommender Systems}}. ACM, \bibinfo{pages}{72--76}.
\newblock


\bibitem[\protect\citeauthoryear{Xu, Wang, Zhang, and Chen}{Xu
  et~al\mbox{.}}{2007}]%
        {xu2007privacy}
\bibfield{author}{\bibinfo{person}{Yabo Xu}, \bibinfo{person}{Ke Wang},
  \bibinfo{person}{Benyu Zhang}, {and} \bibinfo{person}{Zheng Chen}.}
  \bibinfo{year}{2007}\natexlab{}.
\newblock \showarticletitle{Privacy-enhancing personalized web search}. In
  \bibinfo{booktitle}{\emph{Proceedings of the 16th international conference on
  World Wide Web}}. ACM, \bibinfo{pages}{591--600}.
\newblock


\bibitem[\protect\citeauthoryear{Yang, Soboroff, Xiong, Clarke, and
  Garfinkel}{Yang et~al\mbox{.}}{2016}]%
        {yang2016privacy}
\bibfield{author}{\bibinfo{person}{Hui Yang}, \bibinfo{person}{Ian Soboroff},
  \bibinfo{person}{Li Xiong}, \bibinfo{person}{Charles~LA Clarke}, {and}
  \bibinfo{person}{Simson~L Garfinkel}.} \bibinfo{year}{2016}\natexlab{}.
\newblock \showarticletitle{Privacy-preserving ir 2016: Differential privacy,
  search, and social media}. In \bibinfo{booktitle}{\emph{Proceedings of the
  39th International ACM SIGIR conference on Research and Development in
  Information Retrieval}}. \bibinfo{pages}{1247--1248}.
\newblock


\bibitem[\protect\citeauthoryear{Zhang, Yang, and Singh}{Zhang
  et~al\mbox{.}}{2016}]%
        {zhang2016anonymizing}
\bibfield{author}{\bibinfo{person}{Sicong Zhang}, \bibinfo{person}{Hui Yang},
  {and} \bibinfo{person}{Lisa Singh}.} \bibinfo{year}{2016}\natexlab{}.
\newblock \showarticletitle{Anonymizing query logs by differential privacy}. In
  \bibinfo{booktitle}{\emph{Proceedings of the 39th International ACM SIGIR
  conference on Research and Development in Information Retrieval}}.
  \bibinfo{pages}{753--756}.
\newblock


\bibitem[\protect\citeauthoryear{Zhu, Xiong, and Verdery}{Zhu
  et~al\mbox{.}}{2010}]%
        {zhu2010anonymizing}
\bibfield{author}{\bibinfo{person}{Yun Zhu}, \bibinfo{person}{Li Xiong}, {and}
  \bibinfo{person}{Christopher Verdery}.} \bibinfo{year}{2010}\natexlab{}.
\newblock \showarticletitle{Anonymizing user profiles for personalized web
  search}. In \bibinfo{booktitle}{\emph{Proceedings of the 19th international
  conference on World wide web}}. ACM, \bibinfo{pages}{1225--1226}.
\newblock


\bibitem[\protect\citeauthoryear{Zimmerman, Thorpe, Fox, and
  Kruschwitz}{Zimmerman et~al\mbox{.}}{2019}]%
        {zimmerman2019investigating}
\bibfield{author}{\bibinfo{person}{Steven Zimmerman}, \bibinfo{person}{Alistair
  Thorpe}, \bibinfo{person}{Chris Fox}, {and} \bibinfo{person}{Udo
  Kruschwitz}.} \bibinfo{year}{2019}\natexlab{}.
\newblock \showarticletitle{Investigating the Interplay Between Searchers'
  Privacy Concerns and Their Search Behavior}. In
  \bibinfo{booktitle}{\emph{Proceedings of the 42nd International ACM SIGIR
  Conference on Research and Development in Information Retrieval}}.
  \bibinfo{pages}{953--956}.
\newblock


\end{thebibliography}
}

\end{document}